\documentclass{article}
\usepackage{epsf}
\newcommand{\beq}{\begin{equation}}
\newcommand{\eeq}{\end{equation}}
\newcommand{\beqa}{\begin{eqnarray}}
\newcommand{\eeqa}{\end{eqnarray}}

\newcommand{\vs}{\vspace{-0.25cm}}

\begin{document}

\hfill {\tiny FZJ-IKP(Th)-2002-07}

\vspace{3cm}

\begin{center}

{{\Large\bf Analysis of the pion--kaon sigma--term and related topics}}\footnote{Work
  supported in part by Studienstiftung des deutschen Volkes.}

\end{center}

\vspace{.3in}

\begin{center}
{\large Matthias Frink, Bastian Kubis, Ulf-G. Mei{\ss}ner}

\bigskip

\bigskip

Forschungszentrum J\"ulich, Institut f{\"u}r Kernphysik (Theorie)\\
D-52425 J\"ulich, Germany\\
{\it emails: m.frink,b.kubis,u.meissner@fz-juelich.de}\\
\end{center}

\vspace{0.5in}

\thispagestyle{empty}

\begin{abstract}
\noindent We calculate the one--loop contributions to the
difference $\Delta_{\pi K}$ between the isoscalar on--shell pion--kaon 
scattering amplitude at the Cheng--Dashen point  and
the scalar form factor $\Gamma_{K} (2M_\pi^2)$ in the framework of three 
flavor chiral perturbation theory. These corrections turn out to be
small. This is further sharpened by treating the kaons as heavy
fields (two flavor chiral perturbation theory). We also analyze
the two-loop corrections to the kaon scalar form factor based on a dispersive
technique. We find that these corrections are smaller than in the
comparable case of the scalar form factor of the pion. This is related
to the weaker final state interactions in the pion--kaon channel.
\end{abstract}

\vspace{0.5in}

\begin{center}
\end{center}

\vfill

\newpage

\section{Introduction}
\def\theequation{\arabic{section}.\arabic{equation}}
\setcounter{equation}{0}

In QCD, the mass terms for the three light quarks
$u$,$d$, and $s$ can be measured in  the so--called sigma--terms.
These are matrix elements of the scalar quark currents $m_q \bar{q}q$  
in a given hadron $H$, $\langle H| m_q \bar{q}q | H \rangle$, with $H$
e.g.  pions, kaons or nucleons. Since no external scalar probes
are available, the determination of these matrix elements proceeds
by analyzing four--point functions, more precisely Goldstone boson--hadron 
scattering amplitudes in the unphysical region,
$\phi (q) + H(p) \to \phi (q') + H (p')$ (note that the hadron can
also be a Goldstone boson).  The determination of the sigma--terms starts from
the generic  low-energy theorem (such a low--energy theorem
was first formulated for pion--nucleon scattering~\cite{BPP}) 
for the isoscalar scattering  amplitude $A(\nu,t)$
\beq\label{LET}
F^2 A(t,\nu) = \Gamma (t) + q'^\mu q^\nu \, r_{\mu\nu}~,
\eeq
where $F$ is the Goldstone boson decay constant and 
$\Gamma (t)$ is the pertinent scalar form factor
\beq
 \Gamma (t) = \langle H(p') \, | \, {m_q} \bar q q \, | \, H (p) \rangle ~, \quad 
t = (p'-p)^2~,
\eeq
employing  the standard Mandelstam variables $s,t,u$ to describe the 
scattering process, with $s+t+u= 2M_H^2 + 2M_\phi^2$, and further 
introducing the crossing variable $\nu = s-u$.
At zero momentum transfer, this scalar form factor gives the desired sigma--term,
\beq
\Gamma (0) = 2 M_H \, \sigma_{\phi H}~, 
\eeq
for appropriately normalized hadron states (note that sometimes one
uses $M_\phi$ for the normalization). Furthermore, in Eq.(\ref{LET})
$r_{\mu\nu}$ is the so--called remainder, which is not determined by
chiral symmetry. However, it has the same analytical structure as the
scattering amplitude. To determine the sigma--term, one has to work in a
kinematic region where this remainder is small, otherwise a precise determination
is not possible. Beyond tree level, the region where the remainder is small
shrinks to  the so--called Cheng-Dashen (CD) point~\cite{cd}, which e.g. for pion
scattering off other hadrons is given by
\beq
t = 2M_\pi^2 ~, \quad \nu = 0~,
\eeq
which clearly lies outside the physical region for elastic scattering
but well inside the Lehmann ellipse. The most studied reaction to determine
a sigma--term  is certainly elastic
pion--nucleon scattering $\pi N \to \pi N$, 
but  the best understood process is low energy
pion-pion scattering $\pi\pi \to \pi\pi$ (for a beautiful sigma-term
analysis for that case, see
\cite{GS}). Much less is known for processes involving kaons,
in particular for (anti)kaon--nucleon scattering, which is of interest
for particle, nuclear and astrophysics. One of the reasons is the
large kaon mass, which moves the corresponding CD--point to 
$t = 2M_K^2$, far away from
the physical region. That makes the interpolation from the data much more
difficult than in the pion case. In addition, there are open channels below
threshold or even resonances (for $\bar K N \to \bar K N$). There are also
less high precision scattering data. Before addressing these issues, it
is therefore mandatory to understand the simplest process involving
strange quarks, i.e. elastic pion--kaon scattering. This reaction has
attracted much recent interest, see e.g. \cite{AP}, 
mostly triggered by the intended  lifetime measurement of $\pi K$ atoms 
at CERN \cite{DIRAC}, but also as a theoretical laboratory to study the
question whether the kaon can be considered as a heavy particle, see
\cite{Roessl,Oul}. Therefore, as an intermediate step it was proposed 
to analyze the sigma--term in pion--kaon scattering~\cite{GS}. This is done in
this paper in two ways. In section~\ref{sec:oneloop} 
we use standard three flavor chiral perturbation
theory (CHPT)~\cite{GL}, treating the pions and the kaons as (pseudo) Goldstone bosons
of the spontaneously broken chiral symmetry of QCD. We discuss the one-loop 
representation of the scalar pion-kaon form factor and of the isospin-even
$\pi K$ scattering amplitude and deduce the size of the remainder at the CD--point.
In section~\ref{sec:heavy} we analyze the sigma--term in the heavy--kaon framework, 
which helps to  understand the results obtained in SU(3) CHPT. 
To further analyze the stability 
of our results, we calculate in section~\ref{sec:twoloops}
the two-loop corrections to the scalar pion-kaon 
form factor in the threshold region. We employ the dispersive technique 
of Ref.~\cite{GaMe},  despite the fact that the technology for explicit 
two-loop corrections exists, see e.g. \cite{twoloop}. However, for the estimate 
of these effects needed here the use of analyticity and unitarity combined with 
chiral constraints is sufficient.
We end the paper with a short summary in section~\ref{sec:summ}. A related low-energy
theorem for soft kaons is analyzed in appendix~\ref{app:A}.
Further technicalities and definitions are relegated to the following appendices.

\section{Basic considerations}
\def\theequation{\arabic{section}.\arabic{equation}}
\setcounter{equation}{0}
\label{sec:basics}
Elastic pion--kaon scattering can be parameterized
in terms of an isospin~1/2 and an isospin~3/2 amplitude, called 
$T^{1/2}(s,t)$ and $ T^{3/2}(s,t)$, respectively, and $s,t,u$ are
the conventional Mandelstam variables. Note that since these
are subject to the constraint $s+t+u =2 M_\pi^2 + 2M_K^2$, it suffices
to specify two arguments like $s$ and $t$ or $\nu$ and $t$.
Here, we are interested in the isoscalar amplitude
\beq 
T^+_{\pi K} (s,t) \equiv A_{\pi K}(s,t) 
= \frac{1}{3} \, T^{1/2}(s,t) +  \frac{2}{3} \, T^{3/2}(s,t)~. 
\eeq
More precisely, this amplitude can be obtained entirely from the 
isospin~3/2 amplitude because of the  $s \leftrightarrow u$ crossing
relation (for clarity, we exhibit here all three arguments of the scattering
amplitude),
\beq
T^{1/2} (s,t,u) = \frac{3}{2} T^{3/2} (u,t,s) - \frac{1}{2} T^{3/2} (s,t,u)~.
\eeq
Furthermore, the reaction $\pi^+ (q) + K^+ (p) \to  \pi^+ (q') + K^+ (p')$
defines  the isospin~3/2 amplitude,
\beq
\langle\, \pi^+ (q') K^+ (p') \, {\rm out}\, | \,\pi^+ (q)  K^+ (p)\, {\rm in}
\, \rangle = i \, (2\pi)^4 \, \delta^{(4)} (p+q-p'-q') \, T^{3/2} (s,t,u)~.
\eeq
Note also that $T^+$ is even under $s \leftrightarrow u$ crossing, while
the  isovector amplitude $T^- =(T^{1/2}-T^{3/2})/3$ is odd. The partial
wave expansion for the $\pi K$ scattering amplitudes takes the form
\beq
T^I (s,t) = 32 \pi \sum_{l=0}^\infty (2l+1) \, t_l^I (q) \, P_l (z)~,
\eeq
in terms of the squared momentum transfer $t=-2q^2(1-z)$ and the cosine of
the scattering angle, $z = \cos(\theta)$.

\medskip\noindent
The low--energy theorem, Eq.(\ref{LET}), takes the form
\beq\label{CDK}
F^2 \, A_{\pi K}^{\rm CD} = \Gamma_{K} (2M_\pi^2) +
\Delta_{\pi K}~.
\eeq
with $F^2$ expressed in terms of the pion ($F_\pi$) or the kaon $(F_K)$ decay constants 
or the product thereof. From the view point of the chiral expansion, all of
these choices are legitimate. This has most notable consequences for
the remainder because it affects its leading (fourth) order
expression. Therefore,  the fact that $F_\pi \neq F_K$ will play an important
role in the numerical analysis discussed below, related to a particular
chiral SU(2) breaking effect within a three flavor calculation (as explained below).
The pertinent scalar kaon form factor is
\beq\label{GammapiK}
 \Gamma_{K} (t) = \langle K^0(p') \, | \, \hat{m} (\bar u u + 
 \bar d d) \, | \, K^0 (p) \rangle ~, \quad \hat{m} = \frac{1}{2} (m_u+m_d)~.
\eeq
At $t=0$ this defines the $\pi K$ sigma--term,
\beq
2M_\pi \, \sigma_{\pi K} =  \Gamma_{K} (0)~.
\eeq
In what follows, we will analyze the size of the remainder at the CD--point
in the isospin limit $m_u = m_d = \hat{m}$ to one loop accuracy, neglecting
also electromagnetic isospin violation. To get an idea
about possible higher order corrections, we will also calculate the scalar form
factor $\Gamma_{K} (t)$ beyond one loop, following the approach of Ref.~\cite{GaMe}.
In appendix~\ref{app:A}, we analyze a similar low-energy theorem
taking the kaons as soft. 
 
\section{Analysis of \boldmath{$\sigma_{\pi K}$} 
in SU(3) chiral perturbation theory}
\def\theequation{\arabic{section}.\arabic{equation}}
\setcounter{equation}{0}
\label{sec:oneloop}

The tool to systematically calculate low--energy
QCD Green functions and transition currents is chiral perturbation
theory. It amounts to a systematic expansion around the chiral limit
in terms of two small parameters related to the quark masses and the
external momenta \cite{GL}. In the chiral limit of vanishing quark masses,
pions, kaons and etas are massless Goldstone bosons, but in nature
the quark masses are finite, in particular
the strange quark is much heavier than the light up and down quarks, which is reflected
in the difference of the expansion parameters for two and three flavor CHPT,
$M_\pi^2 / (4\pi F_\pi)^2=0.02$ and  $M_K^2 / (4\pi F_\pi)^2 = 0.2$, respectively. 
This large difference is at the heart of the heavy kaon approach to be
discussed below. Here, we analyze the $\pi K$ sigma--term to
the first non--trivial order, i.e to one loop accuracy in the standard
scenario of a large quark condensate, based on the one--loop
representation for  $\pi K$ scattering given in~\cite{BKMpik}. An analysis
of $\pi K$ scattering to leading order in generalized CHPT 
can be found in~\cite{KSSF}. 
Also needed in the analysis of the remainder at the CD-point is the
fourth order representation of the $\pi K$ scalar form factor, first
given explicitly in Ref.~\cite{MO}. It has  the form 
\begin{eqnarray} 
\Gamma_{K}(t)&=&
\frac{M_{\pi}^2}{2} \Bigl\{ 1
+\frac{1}{F^2}\Bigl[L_4^r\bigl(-32M_{K}^2+16
t\bigr)+L_{5}^r\bigl(8M_{\pi}^2-16 M_{K}^2+4t\bigr)+L_6^r 64
M_{K}^2 +L_8^r\bigl(-16M_{\pi}^2+32 M_{K}^2\bigr) \nonumber\\
&-&\frac{1}{2}M_{\pi}^2 \mu_{\pi}+\bigl(-\frac{1}{6}
M_{\pi}^2+\frac{2}{3} M_K^2\bigr)\mu_{\eta}-\frac{3}{4}t J^r_{\pi
  \pi}\bigl(t\bigr) -\frac{3}{4}t J^r_{KK}\bigl(t\bigr)
+\bigl( \frac{2 }{9}M_K^2-\frac{1}{4} t \bigr) J^r_{\eta
  \eta}\bigl(t\bigr)\Bigr] \Bigr\}+{\cal O}(p^6)  ~. \label{gammak}
\end{eqnarray}
Here, the $J_{PQ}^r$ are the renormalized loop functions as defined in \cite{GL} 
and $\lambda$ is the scale of dimensional regularization. We set $\lambda = M_\rho$. 
We use the operator basis of \cite{GL}.
Furthermore,
\beq\label{mudef}
\mu_P = \frac{1}{(4\pi)^2} \ln \frac{M_P^2}{\lambda^2}~, \quad
(P = \pi, K , \eta)~.
\eeq
We remark that setting $F = F_\pi$, the fourth order contribution amounts
to a 22\% correction to the tree level result at the two--pion threshold,
$t = 4M_\pi^2$, which is fairly small for a three flavor observable.
For comparison, the scalar form factor of the pion is affected by
a 29\% correction at the two-pion threshold. We will come back to this
topic in Sec.~\ref{sec:twoloops}. It is also of interest to analyze the
Taylor expansion of $\Gamma_{K}(t)$ around $t = 0$,
\beq
\Gamma_{K}(t) =\Gamma_{K}(0) \, \left( 1 + \frac{1}{6} \langle r^2_S 
\rangle_{K} \, t + {\cal O}(t^2) \, \right)~,
\eeq
in terms of the scalar radius. 
To get a handle on the theoretical uncertainty, we use two sets of values for the 
low--energy constants $L_i$ and their corresponding uncertainties. 
Set~1 is from Ref.\cite{daphne} and set~2 from Ref.\cite{abt} (more
precisely, we use the so-called central fit).
We find
\beq\label{rK}
\langle r^2_S \rangle_{K} = \biggl\{ \begin{tabular}{ll}
        $(0.30 \pm 0.23)$~{\rm fm}$^2$ & $\quad$ {\rm set}~1~, \\
        $(0.38 \pm 0.02)$~{\rm fm}$^2$ & $\quad$ {\rm set}~2~.  
        \end{tabular} 
\eeq
Two remarks are in order. First, the central value is  smaller
than the scalar pion radius, $\langle r^2_S \rangle_{\pi} \simeq 0.6\,$fm$^2$,
pointing towards smaller final state interactions. Second, the uncertainty due
to the LECs is fairly large for set~1 but much smaller for set~2. This
can be traced back that in the second set, the variation in the
OZI--violating LEC $L_4$ is set to zero, $\Delta L_4 =0$, but it is
sizeable for the first set,  $\Delta L_4 \simeq 0.5 \cdot
10^{-3}$. This shows that this observable is very sensitive to this
particular LEC. The large uncertainty due to the variations in the LEC 
also indicates that the chiral logarithms
encoded in the loop contribution play a less distinct role as compared to isoscalar
S-wave pion-pion interactions. We note that the normalization of the
form factor is $\Gamma_K (0) = 0.52 \,(0.53)\, M_\pi^2$ for set 1 (2).

\medskip\noindent
We turn to the remainder at the  CD--point.
To leading order (tree level) it vanishes, as noted before. 
From  the explicit one--loop expressions for the scalar form factor and 
for the $\pi K$ scattering amplitude, it is straightforward to deduce the expression
for the remainder at the CD-point. For
completeness, we give here its explicit form using
the normalization $F^2 = F_\pi^2$, which is natural if one considers
the kaon as the heavy particle (much like a nucleon) from which the
pion scatters. We obtain\footnote{Note that in terms of order $p^4$,
we always set
$F=F_\pi$. This is legitimate to the accuracy we are working.}
\beq\label{CDpiK}
\Delta_{\pi K}^{\rm CD} = \frac{M_\pi^4}{F^2} \left[ L^r_{\rm CD} (\lambda)
+\sum_{P=\pi,K,\eta} \frac{{\cal P}_P}{(4\pi)^2}
\ln\frac{M_P}{\lambda} + {\cal P}_1 \,
J_{\pi K}^r (M_K^2) + {\cal P}_2 \, J_{K\eta}^r (M_K^2)
-\frac{{\cal P}_3}{(4\pi)^2}   \right]~,
\eeq
with $L^r_{\rm CD} (\lambda) = 2(4L_2^r (\lambda) + L_3 - 2L_5^r
(\lambda) + 4L_8^r (\lambda))$ a combination of
low--energy constants. Furthermore,
\beq
\begin{array}{lll}
{\cal P}_\pi = x (x-1) ~,&  {\cal P}_K = \frac{1}{6}(1-2x)~, 
&  {\cal P}_\eta = \frac{1}{9} \left( {x^2} - x -2\right)~, \\
{\cal P}_1 = \frac{1}{2} \left( -{x^2} +\frac{3x}{2} -1\right)~, 
& {\cal P}_2 = \frac{1}{9} \left( 1-\frac{x}{4} - \frac{x^2}{2}\right)~, 
&  {\cal P}_3 = \frac{1}{6} \left( 1 + \frac{x}{2} \right)~,   \\
\end{array}
\eeq
with $x = M_\pi^2 / (2M_K^2) \simeq 1/26$. It is remarkable that no 
terms $\propto M_\pi^2 M_K^2$ appear. This is, of course, different
if one chooses $F^2 = F_\pi F_K$ or $F^2 = F_K^2$ because 
\beq
\frac{F_K}{F_\pi} = 1 + \frac{4 L_5^r}{F_0^2} \, (M_K^2 - M_\pi^2) +
{\rm chiral}~{\rm logs}~,
\eeq
where $F_0$ is the leading term in the quark mass expansion of the
Goldstone boson decay constants. This will reflect itself in 
terms $\sim L_5^r \,M_\pi^2 M_K^2$ in $\Delta_{\pi K}^{\rm CD}$. 
We refrain, however, from giving
the complete analytical formulae for these cases here.

\medskip\noindent Our numerical results for the amplitude, the scalar form factor
and the relative size of the remainder at the CD-point,
$R = \Delta_{\pi K} / (F^2 A_{\pi K}^{\rm CD})$, are collected in table~\ref{tab:rem}. 
For the choice $F = F_\pi^2$, the remainder at the CD--point is very small,
constituting a true SU(2) result as explained below. Even for the  choice
of $F^2 =F_\pi F_K$, the resulting numbers are still on the low side expected 
from SU(3) breaking $\sim (M_K/\Lambda_\chi)^2
\simeq 0.2$ (with $\Lambda_\chi = 4\pi F_\pi$). 
This observation also holds individually for the scattering amplitude and for
the form factor at the CD--point. These results are comparable to what is found in the
analysis of the pion sigma--term~\cite{GS}. If one normalizes $F^2 A_{\pi K}^{\rm CD}$
and  $\Gamma_{\pi K} (2M_\pi^2)$ at tree level to one 
(for an easier comparison with the pion case, 
see below), the first row of table~\ref{tab:rem}
reads
\beq
\begin{tabular}{ccccc}
1.14 & = & 1.10 & + & 0.04\phantom{~,} \\
$F^2 A_{\pi K}^{\rm CD}$ & = & $\Gamma_{K} (2M_\pi^2)$  & + & $\Delta_{\pi K}$~,\\
\end{tabular}
\eeq
astonishingly close to the tree level result.  The 
remainder amounts to a correction of $0.04M_\pi/2 \simeq 2.8\,$MeV.
It is instructive to give the results for the pion case~\cite{GS}
\beq
\begin{tabular}{ccccc}
1.14 & = & 1.09 & + & 0.05\phantom{~,} \\
$F^2_\pi A_{\pi}^{\rm CD}$ & = & $\Gamma_{\pi} (2M_\pi^2)$  & + & $\Delta_{\pi}$~.\\
\end{tabular}
\eeq
We note that the remainder is comparable
to the pion-kaon case, for the pion--sigma term it amounts to a correction
of about 3.5$\,$MeV.
\renewcommand{\arraystretch}{1.2}
\begin{table}[hbt]
\begin{center}
\begin{tabular}{|l|c|c|c|c|}
    \hline
$F^2$       & $L_i$ set 
            &  $F^2 A_{\pi K}^{\rm CD}$ [$M_\pi^2$]
            & $\Gamma_{K}   (2M_\pi^2)$ [$M_\pi^2$] & $R$ [\%] \\
    \hline
$F_\pi^2$   & 1 & $0.572 $   &  $0.551 $    &  $3.7$     \\
$F_\pi F_K$ & 1 & $0.679 $   &  $0.551 $    & $18.9$     \\
$F_K^2$     & 1 & $0.785 $   &  $0.551 $    & $29.8$     \\
$F_\pi^2$   & 2 & $0.600 $   &  $0.587 $    &  $2.2$     \\
$F_\pi F_K$ & 2 & $0.681 $   &  $0.587 $    & $13.8$     \\
$F_K^2$     & 2 & $0.762 $   &  $0.587 $    & $23.0$     \\
\hline\hline
\end{tabular}
\centerline{\parbox{11cm}
{\caption{Size of the remainder at the CD--point for various choices of the
          meson decay constants and the low--energy constants $L_i^r (M_\rho)$.
         \label{tab:rem}}}}
\end{center} 
\end{table}
\medskip 

\noindent
To further illustrate the situation  we consider so-called scale relations.
This amounts to an expansion of both the scalar form factor and the
$\pi K$ amplitude at the CD--point in powers of $M_{\pi}$ and representing 
the occurring terms in terms of the chiral logarithms $  \ln ({
  M_P^2}/{\Lambda^2})\,$, 
$P \in \{\pi, K, \eta\}\,$, via appropriately defined scales
$\Lambda_i\,$. 
The results are only given for the $1/F_{\pi}^2$ choice of
normalization. Consider first
$\Gamma_{K}$ at the CD--point. The non--logarithmic terms in 
the coefficients of $M_{\pi}^4$ respectively $M_{\pi}^2 M_K^2\,$, 
i.e. constant terms and the contributions proportional to the LECs, 
are absorbed into common scales $\Lambda_{1/2}$. The so defined scales
$\Lambda_{1/2}$ are unique and independent of the meson masses. 
We find:
\begin{eqnarray}
\Gamma_{K} ( 2 M_{\pi}^2 )
&=& \frac{M_{\pi}^2}{2}+\frac{M_\pi^2}{(4\pi F)^2}\,
\biggl[M_{\pi}^2 \Bigl(- \ln \bigl( \frac{ M_{\pi}^2}{\Lambda_1^2}\bigr)
-\frac{3}{4}\ln \bigl( \frac{ M_{K}^2}{\Lambda_1^2}\bigr)
-\frac{1}{3}\ln \bigl( \frac{ M_{\eta}^2}{\Lambda_1^2}\bigr) \Bigr)
+\frac{4}{9} M_K^2\ln \bigl( \frac{ M_{\eta}^2}{\Lambda_2^2}
\bigr)\nonumber \\ 
& &\quad\qquad\quad\quad\qquad\quad +\frac{217}{720}\frac{M_{\pi}^4}{M_K^2}
+\frac{1417}{20160}\frac{M_{\pi}^6}{M_K^4} \biggr] 
+{\cal O} \biggl(\frac{M_{\pi}^{10}}{F^2 M_K^6} \biggr)~,
 \end{eqnarray}
where 
\begin{eqnarray}
\Lambda_1 &=& \lambda \exp \biggl[\frac{6}{25} \Bigl( \bigl( 4\pi\bigr)^2 
\bigl(16 L_4^r+8L_5^r-8L_8^r\bigr)-\frac{3}{8}\pi -\frac{5}{18} \Bigr)\biggr]
= 527~ {\rm MeV}~, \nonumber \\
\Lambda_2 &=& \lambda \exp \biggl[\frac{9}{8} 
\Bigl( \bigl( 4\pi\bigr)^2 \bigl(16 L_4^r+8L_5^r-32 L_6^r-16L_8^r\bigr)
-\frac{1}{9}  \Bigr)\biggr] = 511~{\rm MeV}~,
\end{eqnarray}
employing the LEC values of set~1. Note that these expressions are
independent of the regularization scale $\lambda$ since the logarithmic
scale dependence of the $L_i$ cancels the explicit factor of $\lambda$.
The corresponding expression for 
$ F_{\pi}^2 A^{\rm CD}_{\pi K}$ reads: 
\begin{eqnarray}
F_{\pi}^2 A^{\rm CD}_{\pi K}
&=&\frac{M_{\pi}^2}{2}+\frac{M_\pi^2}{(4\pi F)^2}\,
\biggl[M_{\pi}^2\Bigl(- \ln \bigl( \frac{ M_{\pi}^2}{\tilde{\Lambda}_1^2}
\bigr)-\frac{9}{8}\ln \bigl( \frac{ M_{K}^2}{\tilde{\Lambda}_1^2}\bigr)
-\frac{3}{8}\ln \bigl( \frac{ M_{\eta}^2}{\tilde{\Lambda}_1^2}\bigr) \Bigr)
+\frac{4}{9} M_K^2\ln \Bigl( \frac{ M_{\eta}^2}{\tilde{\Lambda}_2^2}
\Bigr)-\frac{\pi}{2} \frac{ M_{\pi}^3}{M_K}
\nonumber \\ 
&+&\frac{M_{\pi}^4}{M_K^2}\Bigl(-\frac{1}{2}\ln \bigl( 
\frac{ M_{\pi}^2}{\lambda^2}\bigr)
+\frac{35}{64}\ln \bigl( \frac{ M_{K}^2}{\lambda^2}\bigr)
-\frac{3}{64}\ln \bigl( \frac{ M_{\eta}^2}{\lambda^2}\bigr)
+\frac{901}{2160}
-\frac{1}{36} \sqrt{2}\arctan \bigl( \sqrt{2}\bigr)\Bigr)
+\frac{7 \pi}{16} \frac{ M_{\pi}^5}{M_K^3}
\nonumber \\
&+&\frac{M_{\pi}^6}{M_K^4}
\Bigl(\frac{5}{16}\ln \bigl( \frac{ M_{\pi}^2}{\lambda^2}\bigr) 
-\frac{81}{256}\ln \bigl( \frac{ M_{K}^2}{\lambda^2}\bigr)
+\frac{1}{256}\ln \bigl( \frac{ M_{\eta}^2}{\lambda^2}\bigr)
-\frac{3181}{15120} -\frac{23}{1152} \sqrt{2}\arctan 
\bigl( \sqrt{2}\bigr)\Bigr) \biggr] \nonumber \\
&& \qquad\qquad\qquad\qquad\qquad \qquad\qquad\qquad\qquad\qquad
\qquad\qquad\qquad\qquad\qquad
+{\cal O} \biggl(\frac{M_{\pi}^{10}}{F^2 M_K^6} \biggr)~,
\end{eqnarray}
and
\begin{eqnarray}
\tilde{\Lambda}_1 &=& \lambda \exp \biggl[\frac{1}{5} \biggl( 
\Bigl( 4\pi\Bigr)^2 \Bigl(8 L_2^r+2L_3^r+16L_4^r+4 L_5^r\Bigr)
-\frac{3}{8}\pi-\frac{1}{18}
+\frac{4}{27} \sqrt{2}\arctan \bigl( 
\sqrt{2}\bigr)\biggr)\biggr]=724~{\rm MeV}~, \nonumber \\
\tilde{\Lambda}_2 &=& \lambda \exp \biggl[\frac{9}{8} \Bigl( 
\bigl( 4\pi\bigr)^2 \bigl(16 L_4^r+8L_5^r-32 L_6^r-16L_8^r\bigr)
-\frac{1}{9}  \Bigr)\biggr]=511~{\rm MeV}~.
\end{eqnarray}
For the terms of order ${\cal O}({M_{\pi}^6}/{F^2 M_K^2})$ 
and ${\cal O}({M_{\pi}^8}/{F^2 M_K^4})$ it is not possible to 
incorporate the constants in the coefficients into 
a universal scale  in each of the chiral logarithms, 
as can be traced back to their origin in the regularization procedure. 
Chiral logarithms never appear alone, their occurrence is, 
moreover, accompanied by a pole term $L\,$. In the absence of 
counterterms proportional to $M_{\pi}^6/(F^2 M_K^2)$ or $ M_{\pi}^8/
(F^2 M_K^4)\, $, 
renormalizability requires the coefficients of $L$ to add up to zero 
for a given order, and the same is therefore true for the coefficients 
of the logarithms. It is thus impossible to allocate to each 
chiral logarithm its proportionate fraction of the non--logarithmic 
contributions, and neither is it then possible to define a common scale. 
For the purpose of illustrating the similarity structure 
of the scalar form factor and of the scattering amplitude 
we analyze the scales in those terms which behave as $M_{\pi}^4$ 
and $ M_{\pi}^2 M_K^2\,$. With $\Lambda_2$ and $\tilde{\Lambda}_2$ 
close to the eta mass it is clear that the potentially large corrections 
involving $M_K^2$ are individually small. Since both scales are 
even identical, these contributions cancel completely in the remainder. 
In the case of $\Lambda_1$ and $\tilde{\Lambda}_1$ cancellations are 
not complete, as $\tilde{\Lambda}_1$ is larger than $\Lambda_1\,$, 
whose value is again not much different from the eta mass.

\section{Analysis of  \boldmath{$\sigma_{\pi K}$} in the heavy--kaon approach}
\label{sec:heavy}
\def\theequation{\arabic{section}.\arabic{equation}}
\setcounter{equation}{0}

So far, we have considered the kaons and the pions on equal footing, namely
as pseudo-Goldstone bosons of the spontaneously broken chiral symmetry of QCD,
with their finite masses related to the non-vanishing current quark masses.
However, the fact that the kaons (and also the eta) are much heavier than the
pions might raise the question whether a perturbative treatment in the
strange quark mass is justified. In fact, one can take a very different view 
and consider only the pions as light with the kaons behaving as heavy sources, 
much like a conventional matter field in baryon CHPT. This point of view was 
first considered in the Skyrme model \cite{CK} and has been reformulated in the context of
heavy--kaon chiral perturbation theory (HKCHPT) in Ref.\cite{Roessl} (a closely related
work applying reparameterization invariance instead of the reduction of 
relativistic amplitudes was presented in \cite{Oul}.). Since the kaons appear
now as matter fields, the chiral Lagrangian for pion-kaon interaction decomposes
into a string of terms with a fixed number of kaon fields, 
that is into sectors with $n$ ($n \geq 0$) 
in-coming and $n$ out-going kaons. Here, we consider processes with at most one
kaon in the in/out states. Obviously, the power counting has to be modified due
to the new large mass scale, $M_K$, and as it is the case for baryons, terms
with an odd number of derivatives are allowed. For keeping the paper
self-contained, we give in appendix~\ref{app:heavyK} a more detailed discussion of
the heavy--kaon formulation, following essentially Ref.~\cite{Roessl}.
This appraoch is particularly 
suited to analyze chiral SU(2) theorems for three flavor observables, and it
is therefore natural to reconsider pion-kaon scattering and the issues related
to it discussed in the previous sections\footnote{Note that the general derivation of
the $\pi K$ scattering amplitude  was already done in Ref.~\cite{Roessl}.}.
The heavy kaon formulation can be connected to the standard SU(3)
CHPT approach  by so-called matching relations, which will
be discussed in some detail below for the case of the scalar form factor $\Gamma_K$.
In general, this need not be done, but for practical purposes it
cannot be avoided, there are simply not enough precise low--energy
data to pin down the heavy kaon LECs independently.
The corresponding heavy--kaon Lagrangian for doing this matching is displayed 
in  appendix~\ref{app:heavyKL}.

\medskip
\noindent
Let us first consider the scalar form factor of the kaon. The calculation proceeds
as in the standard case, only that we have to consider new vertices and the loops
are entirely pionic ones. We will again work with the physical masses 
and employ dimensional regularization exactly. To one-loop accuracy, one finds
the following renormalized (finite) representation for $\Gamma_K$:
\begin{eqnarray}
\Gamma_K (t) &=& \frac{M_{\pi}^2}{2} \biggl[8\Bigl((A_3^r+2 A_4^r)
\Bigl(-\frac{1}{2M_\pi^2} + 2C_5^r + 4C_6^r +\frac{1}{F^2} l_3^r \Bigr) 
- 4 C_{13}^r- 4 C_{14}^r- 8 C_{15}^r 
- \frac{1}{8(4\pi F)^2}A_2^r M_K^2 \Bigr) M_{\pi}^2 \nonumber \\ 
&+& \Bigl( 2C_5^r +4C_6^r + \frac{1}{(4\pi F)^2}\frac{A_2^r}{6} M_K^2\Bigr)t
+\frac{1}{F^2}\Bigl(8A_3^r+16 A_4^r+3A_1^r
+\frac{A_2^r}{2}M_K^2 \Bigr) M_{\pi}^2 \mu_{\pi} 
\nonumber \\
&+&\frac{1}{F^2}\Bigl( \bigl(3A_1^r+A_2^r M_K^2+6A_3^r+12 A_4^r\bigr)M_{\pi}^2
 +\bigl(-\frac{3}{2} A_1^r-\frac{1}{4}A_2^r M_K^2\bigr)t\Bigr) 
J^r_{\pi \pi}(t) \biggr] +{\cal O}(p^6)~.
\label{hkgammak}
\end{eqnarray}
In this and in all following formulae, one has $F = F_\pi$.
To provide HKCHPT with predictive power we need the numerical values 
of the renormalized LECs characteristic of the heavy kaon theory, more specifically the
$A_i^r$ and $C_i^r$ appearing in Eq.~(\ref{hkgammak}). 
In principle, these can be obtained from experimental data, in complete analogy to the
determination of  the $L_i\,$ in conventional SU(3) CHPT. In fact, one can translate 
knowledge of the $L_i$ into the heavy kaon theory and thus infer information about the HKCHPT 
parameters. The major difference in both approaches is the treatment of the strange quark 
mass.  While  in SU(3) CHPT $m_s$ serves as an expansion parameter of the chiral series, 
in the heavy  kaon approach $m_s$ does enter as part of the static kaon mass, yet, 
when involved in loops, the kaon is rather dealt with as a heavy quark, i.e. its effects are 
absorbed into the numerical  values of the constants present in any expansion. Having 
calculated an observable quantity in both schemes, a comparison of the two power series 
then yields an expansion of certain combinations 
of HKCHPT parameters in powers of $m_s\,$, where the SU(3) CHPT LECs are incorporated 
into the coefficients. One can thus carry out an order by order investigation as to the role 
which the strange quark mass plays for the heavy kaon constants, and as for their dependence on the 
renormalization scale. This procedure is referred to as matching. The matching relations of a 
sufficient number of observables then provide enough relations among the parameters to solve 
for each of them separately. 
We now return to the scalar form factor of the kaon. For the matching procedure,
we have to bring $\Gamma_K$ from Eq.~(\ref{gammak}) into a form which allows for a direct
comparison with its heavy kaon counterpart, Eq.~(\ref{hkgammak}). 
We rewrite the form factor in terms of $\bar{M}_K\,$ (see appendix~\ref{app:heavyK} for
definition of this and related quantities) and expand in powers  of the light quark mass 
$\hat{m}$ and  the squared momentum transfer $t\,$. Using Eqs.~(\ref{HKloops}), we find
\begin{eqnarray}
\Gamma_K(t) &=& \frac{M_{\pi}^2}{2}+\frac{M_{\pi}^2}{2F^2} \biggl[\bar{M}_K^2\Bigl(-32 L_4^r-16 
L_5^r+64 L_6^r+32 L_8^r+\frac{2}{9} \frac{1}{(4\pi)^2}
+\frac{8}{9} \frac{1}{(4\pi)^2} \ln \bigl(\frac{4}{3} 
\frac{\bar{M}_K^2}{\lambda^2} \bigr)\Bigr)
\nonumber \\
 & & 
+M_{\pi}^2 \Bigl(-16 L_4^r+32 L_6^r+\frac{1}{3} 
\frac{1}{(4\pi)^2}
+\frac{5}{18} \frac{1}{(4\pi)^2} \ln \bigl(\frac{4}{3} 
\frac{\bar{M}_K^2}{\lambda^2} \bigr)\Bigr) -\frac{1}{2}M_{\pi}^2 \mu_{\pi}+t 
\Bigl(16 L_4^r+4 L_5^r-\frac{37}{36} \frac{1}{(4\pi)^2}\nonumber \\ 
& & \quad\quad\quad-\frac{3}{4} \frac{1}{(4\pi)^2} \ln \bigl(\frac{\bar{M}_{K}^2}{\lambda^2} \bigr)
-\frac{1}{4} \frac{1}{(4\pi)^2} \ln \bigl(\frac{4}{3} \frac{\bar{M}_K^2}{\lambda^2} \bigr)\Bigr)
-\frac{3}{4}\frac{t}{F^2} J_{\pi \pi}^r \bigl(t\bigr)\Bigr) \biggr]~.
\label{sffex}
\end{eqnarray}
We then equate the coefficients of the various terms in Eq.~(\ref{sffex}) with the ones
in Eq.~(\ref{hkgammak}) and arrive at the desired matching relations:
\beqa\label{unub}
-4A_3^r-8 A_4^r -1
 &=&\frac{\bar{M}_K^2}{F^2}\Bigl(-32 L_4^r-16 L_5^r+64 L_6^r+32 L_8^r 
+\frac{2}{9} \frac{1}{(4\pi)^2} +\frac{8}{9} \frac{1}{(4\pi)^2} 
\ln \bigl(\frac{4}{3} \frac{\bar{M}_K^2}{\lambda^2} \bigr)\Bigr)\nonumber \\
&& \qquad\qquad\qquad\qquad\qquad \qquad\qquad\qquad\qquad\qquad 
+{\cal O}\Bigl(\bar{M}_K^4\Bigr)~,
\\
\label{unub1}
16 A_3^r C_5^r+32 A_3^rC_6^r + 32 A_4^rC_5^r\!\!\!\! &+&\!\!\!\! 64 A_4^rC_6^r+\frac{8}{F^2}A_3^r 
l_3^r+\frac{16}{F^2} A_4^r l_3^r-32 C_{13}^r-32 C_{14}^r
 -64 C_{15}^r-\frac{1}{F^2} \frac{1}{(4\pi)^2}A_2^r M_K^2 \nonumber \\ 
&=&\frac{1}{F^2}\Bigl(-16 L_4^r+32 L_6^r+\frac{1}{3} 
\frac{1}{(4\pi)^2}+\frac{5}{18} 
\frac{1}{(4\pi)^2} \ln \bigl(\frac{4}{3} \frac{\bar{M}_K^2}{\lambda^2} \bigr)\Bigr)+{\cal O}
\Bigl(\bar{M}_K^2\Bigr)~,
\\
8A_3^r+16 A_4^r+3A_1^r +\frac{A_2^r}{2}M_K^2 
&=& -\frac{1}{2}+{\cal O}\Bigl(\bar{M}_K^2\Bigr)~,
\\
2C_5^r +4C_6^r + \frac{1}{\Lambda_\chi^2} \frac{A_2^r}{6} M_K^2
&=&\frac{1}{F^2}\Bigl(16 L_4^r+4 L_5^r-\frac{37}{36} \frac{1}{(4\pi)^2}-\frac{3}{4} 
\frac{1}{(4\pi)^2} \ln \bigl(\frac{\bar{M}_{K}^2}{\lambda^2} \bigr)-\frac{1}{4} 
\frac{1}{(4\pi)^2} \ln \bigl(\frac{4}{3} \frac{\bar{M}_K^2}{\lambda^2} \bigr)\Bigr)
\nonumber \\
&& \qquad\qquad\qquad\qquad\qquad \qquad\qquad\qquad\qquad\qquad 
+{\cal O}\Bigl(\bar{M}_K^4\Bigr)~,
\label{neumatch} \\
-\frac{3}{2} A_1^r-\frac{1}{4} A_2^r M_K^2
&=&-\frac{3}{4}+{\cal O}\Bigl(\bar{M}_K^2\Bigr)~,
\\
6A_3^r+12 A_4^r+3A_1^r+A_2^r M_K^2 &=&{\cal O}\Bigl(\bar{M}_K^2\Bigr)~.
\eeqa
These conditions, except one, have previously been obtained \cite{Roessl} (note that
we have corrected for some obvious misprints in that paper).
Eq.~(\ref{neumatch}) provides new information, whose source is essentially the 
$t$--dependence of $\Gamma_K\,$. To the accuracy we are working we can neglect the 
term proportional to the kaon mass, such that the essential information from matching 
the scalar form factor of the kaon is given by the following equation:
\begin{equation}
C_5^r+2C_6^r=\frac{1}{F^2} \Bigl(8L_4^r+2L_5^r-\frac{37}{72} \frac{1}{(4\pi)^2}-\frac{3}{8} 
\frac{1}{(4\pi)^2} \ln\bigl(\frac{\bar{M}_{K}^2}{\lambda^2} \bigr)-\frac{1}{8} \frac{1}{(4\pi)^2} 
\ln\bigl(\frac{4}{3}\frac{\bar{M}_{K}^2}{\lambda^2} \bigr)\Bigr)+{\cal O}\Bigl(\bar{M}_K^2\Bigr)~.
\end{equation}
This can, in turn, be used to isolate $C_{13}^r+C_{14}^r+2C_{15}^r$ from (\ref{unub1}) 
with the following result:
\beq
C_{13}^r+C_{14}^r+2C_{15}^r = \frac{1}{F^2} \Bigl(-2L_6^r-\frac{1}{2}L_8^r+\frac{1}{18} 
\frac{1}{(4\pi)^2}+\frac{3}{64} \frac{1}{(4\pi)^2} \ln\bigl(\frac{\bar{M}_{K}^2}{\lambda^2} 
\bigr) +\frac{5}{576} \frac{1}{(4\pi)^2} \ln\bigl(\frac{4}{3}\frac{\bar{M}_{K}^2}{\lambda^2} 
\bigr)\Bigr)+{\cal O}(\bar{M}_K^2)~.
\eeq
As we did in the preceeding section, we handle the theoretical uncertainties by working 
with two sets of values for the SU(3) CHPT LECs. The results are displayed in 
table~\ref{coval}. They reflect the LECs of the heavy kaon theory for
a certain  renormalization 
scale, which is inherited from the standard LECs used in the calculation, i.e. 
$\lambda = M_{\rho}$. We note while the dimension two heavy LEC
combination is well 
determined,
there is a large variation in the dimension four heavy kaon LEC combinations for the two
sets of $L_i$. This is interesting because to a certain extent it
reflects the dependence on the
OZI-violating LECs $L_4^r$ and $L_6^r$.
Employing these matching conditions, the scalar form factor can be
studied numerically. First, we find that the normalization $\Gamma_K
(0)$ increases as compared to the standard SU(3) CHPT case,
$\Gamma_K (0) = 0.56 \, (0.61)\,M_\pi^2$ for set~1~(2) (see also the
discussion below). Second, as a consequence of that,
the corresponding radius shrinks a bit,
\beq\label{rKHK}
\langle r^2_S \rangle_{K} = 0.23 \,\, (0.26)~{\rm fm}^2~, \quad {\rm set}~1\,\,(2)~,
\eeq
compare Eq.~(\ref{rK}). We refrain from repeating the analysis of the 
theoretical error due to the uncertainty in the heavy kaon LECs since this
would only reflect the uncertainty of the $L_i^r$ already discussed in the
preceeding section. At first, the closeness of the values for the scalar
radius using sets 1 or 2 seems puzzling since in the polynomial part of
Eq.~(\ref{hkgammak}) the term linear in $t$ is multiplied by $C_5^r + 2C_6^r$,
which is very different for the two sets. However, these LECs are
very small  and furthermore, this effect is to a large
portion cancelled by the contribution from the term 
$\sim A_1^r  \, J^r_{\pi\pi} (t)$.

\begin{table}[hbt]
\begin{center}
\begin{tabular}{|l||c|c|}
    \hline
  &  set~1  & set 2 \\ \hline
  $A_1^r$                               & $\,\,\,0.68$  & $\,\,\,\,0.52$ \\
  $A_3^r+2A_4^r $                       & $-0.26 $      &  $-0.28 $ \\
  $A_2^r$ [GeV$^{-2}$]                  & $-6.35$       &  $-4.68$  \\
  $B_1^r$ [GeV$^{-2}$]                  & $\,\,\,0.93$  &  $\,\,\,0.56$  \\
  $B_3^r$ [GeV$^{-2}$]                  & $\,\,\,0.83$  &  $\,\,\,\,0.68$  \\
  $C_1^r$ [GeV$^{-2}$]                  & $-1.96$       &  $-0.74$  \\
  $C_5^r+2C_6^r+4C_7^r+2C_8^r+4C_9^r$ [GeV$^{-2}$]
                                        & $-2.03$       &  $-2.03$  \\
  $8(C_{13}^r+C_{14}^r+2C_{15}^r+C_{16}^r)+C_8^r+2C_9^r$ [GeV$^{-2}$]
                                        & $-0.84$       &  $-1.07$  \\
  \hline
  $C_5^r+2C_6^r$ [GeV$^{-2}$]           & $-0.02$       &  $\,\,\,0.15$  \\
  $C_{13}^r+C_{14}^r+2C_{15}^r$[GeV$^{-2}$] & $-0.001$  & $-0.03$  \\
  \hline
  $2C_7^r+C_8^r+2C_9^r$ [GeV$^{-2}$]    & $-1.01$ & $-1.09$ \\
  $8C_{16}^r+C_8^r+2C_9^r$ [GeV$^{-2}$] & $-0.83$ & $-0.82$
  \\
\hline\hline
\end{tabular}
\centerline{\parbox{12cm}
{\caption{Values of some combinations of HKCHPT LECs for various
    choices of the SU(3) CHPT LECs $L_i^r (M_\rho)$.
    The first eight entries are derived from matching the $\pi K$ 
    scattering amplitude (some of these are also found in the analysis
    of the scalar kaon form factor as explained in the text). The next two
    stem from the momentum dependence of $\Gamma_K (t)$. The large variation
    for these two can be traced back to the rather different input values
    for some of the OZI--violating LECs in sets 1 and 2, respectively. The 
    last two are particular combinations  of dimension three LECs which can be 
    obtained from the former relations.
   \label{coval}}}}
\end{center} 
\end{table}

\begin{figure}[htb]
\centerline{
\epsfysize=1.2in
\epsffile{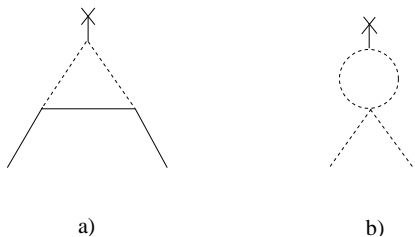}
}
\begin{center}
\parbox{10cm}{\caption{\protect Dominant loop contributions to the
scalar form factor. a) In the pion-nucleon case, this stems from the so--called
triangle graph. Solid (dashed) lines denote nucleons (pions). b)
In the pion--kaon case, one has tadpole like contribution and others.
Here, the dashed lines denote Goldstone bosons.
\label{fig:triangle}}}
\end{center}
\end{figure}

\medskip\noindent
Next, we consider the $\pi K$ amplitude and the remainder at the CD-point. First,
we note that the heavy kaon scattering amplitude has been first evaluated and analyzed in
\cite{Roessl}. However, the amplitude given in that paper is not free of errors,
therefore we give the corrected form in appendix~\ref{app:heavyTpiK}. With that, the
reported discrepancy \cite{Roessl} between the chiral prediction for some of the
threshold parameters in the relativistic and the heavy kaon scheme
disappears. We have also rederived the matching relations from the
amplitude, which mostly agree with the ones in \cite{Roessl}. In two
relations, we found a discrepancy, the corrected formulae are
displayed in  appendix~\ref{app:heavyTpiK}. The numerical results are 
collected in table~\ref{coval}.
Putting pieces together, we arrive at the remainder at the CD-point,
\beq
\Delta_{\pi K} = M_\pi^4 \left( -{A_2^r\over 4} - 4(A_3^r +
  2A_4^r)(C_5^r + 2C_6^r) + 4C_7^r  -16C_{16}^r \right)~.
\eeq 
Note that all non--polynomial pieces have disappeared. This is
consistent with the previous finding because for the choice $F^2 =
F_\pi^2$ in the standard  scenario we had no contribution from 
pure pion loops $\sim J_{\pi \pi}^r$ and the logarithmic terms $\ln
M_\pi$ in the light kaon case, see Eq.~(\ref{CDpiK}), 
only appear at higher orders in the heavy kaon power counting.
Employing the matching relations, we can analyze the LET, Eq.~(\ref{CDK}),
and find (again normalizing the tree result to one)
\beq
\begin{tabular}{ccccc}
1.22 (1.27) & = & 1.18 (1.26) & + & 0.048 (0.014)\phantom{~,} \\
$F^2 A_{\pi K}^{\rm CD}$ & = & $\Gamma_{K} (2M_\pi^2)$  & + & $\Delta_{\pi K}$~,\\
\end{tabular}
\eeq
which means that the relative size of the remainder is 3.9\% 
(1.6\%) for set~1~(2). This is similar to the
results in standard SU(3) CHPT for the choice $F^2 = F_\pi^2$. We also
note that the normalization of the form factor $\Gamma_K (0)$ has
somewhat increased in the heavy kaon approach. This is due
to the value of $A_3^r + 2A_4^r$ which via the matching condition
subsumes some higher order corrections. A similar statement can be 
made for the pion--kaon amplitude. At first sight, this might appear
worrysome but it can be traced back to our treatment of the matching 
conditions, on which we imposed a strict power counting in $\bar{M}_K$.
It would of course also be allowed to include such higher order terms
in the matching conditions.
This would lead to a reduction of the apparent discrepancy between the
heavy kaon and the standard formulation. However, our intention in using the
heavy kaon formulation was not to reproduce exactly the numbers obtained
in the standard case but rather to consider the same observables in a
scheme which treats the kaons very differently. Furthermore, these small isoscalar
observables are also subject to the largest theoretical uncertainties,
a situation similar to the case of pion--nucleon scattering. However,
it is also important to discuss the difference to the
pion--nucleon scattering amplitude. So far, we have stressed the similiarity
between $\pi K$ and $\pi N$ scattering, but there are some differences
due to the absence of three-Goldstone-boson couplings. In the context discussed 
here, this has a major influence on the momentum dependence of the scalar
form factor respectively on the $t$--dependence of the scattering amplitude.
In the pion--nucleon case, the very strong momentum dependence around
$t = 4M_\pi^2$ is due to the fact that the so--called triangle diagram
(see Fig.~\ref{fig:triangle}a) has a singularity on the second Riemann sheet
at $t_c = 4M_\pi^2 - M_\pi^4/m^2 = 3.98 M_\pi^2$, i.e. very close to the
threshold. In fact, in the heavy fermion limit, this singularity coalesces
with the threshold and thus distorts the analytical structure. Such an 
effect can also be seen in the spectral functions of the isovector nucleon
form factors. Quite differently, the $t$--dependence for the pion--kaon
case is given by loop graphs (as shown in Fig.~\ref{fig:triangle}b) and contact
terms that do not have such close--by singularities.  

\medskip\noindent
We end this section by remarking that one may also try to fix the 
heavy kaon LECs $A_i^r,B_i^r$ and $C_i^r$  directly from
a systematic analysis of low--energy data involving kaons. Given however
the scarcity of precise data for processes with a conserved kaon number,
we refrain from performing such an analysis here.

\section{Two-loop representation of {\boldmath $\Gamma_{K}$} 
\def\theequation{\arabic{section}.\arabic{equation}}
\setcounter{equation}{0}
\label{sec:twoloops}}

In sec.~\ref{sec:oneloop} we have discussed the one-loop representation of the scalar 
form factor $\Gamma_K$ of the kaon in the framework of SU(3) CHPT. 
Numerically, the fourth order contributions were found to be 
about $10$ \% at the CD--point $t=2 M_{\pi}^2$ and $22$ \% at the 
two--pion threshold $s=4 M_{\pi}^2\,$. We had also shown that the small
correction to the LET Eq.~(\ref{CDK}) 
was due to the suppression of terms proportional
to powers of the kaon mass when the field normalization $\sim 1/F_\pi^2$
was chosen. Still, in view of possible large higher order corrections
in the S-wave isospin zero channel and the expected slow convergence 
behavior of three flavor chiral perturbation theory, it is mandatory
to estimate the two-loop contributions.
Because two-loop diagrams are awkward to calculate, we 
seek to obtain information concerning contributions of sixth chiral order at 
lower cost by following a different strategy, which is based on the 
unitarity properties of the scattering operator and on the analyticity properties 
of the perturbation series representing $\Gamma_K\,$. We follow essentially
the work of Ref.~\cite{GaMe} on the dispersive representation of the pion
form factors. We do not perform a very precise determination of the occuring
subtraction constants. For our purpose, however, this procedure is of 
sufficient accuracy.

\begin{figure}[htb]
\centerline{
\epsfysize=2.0in
\epsffile{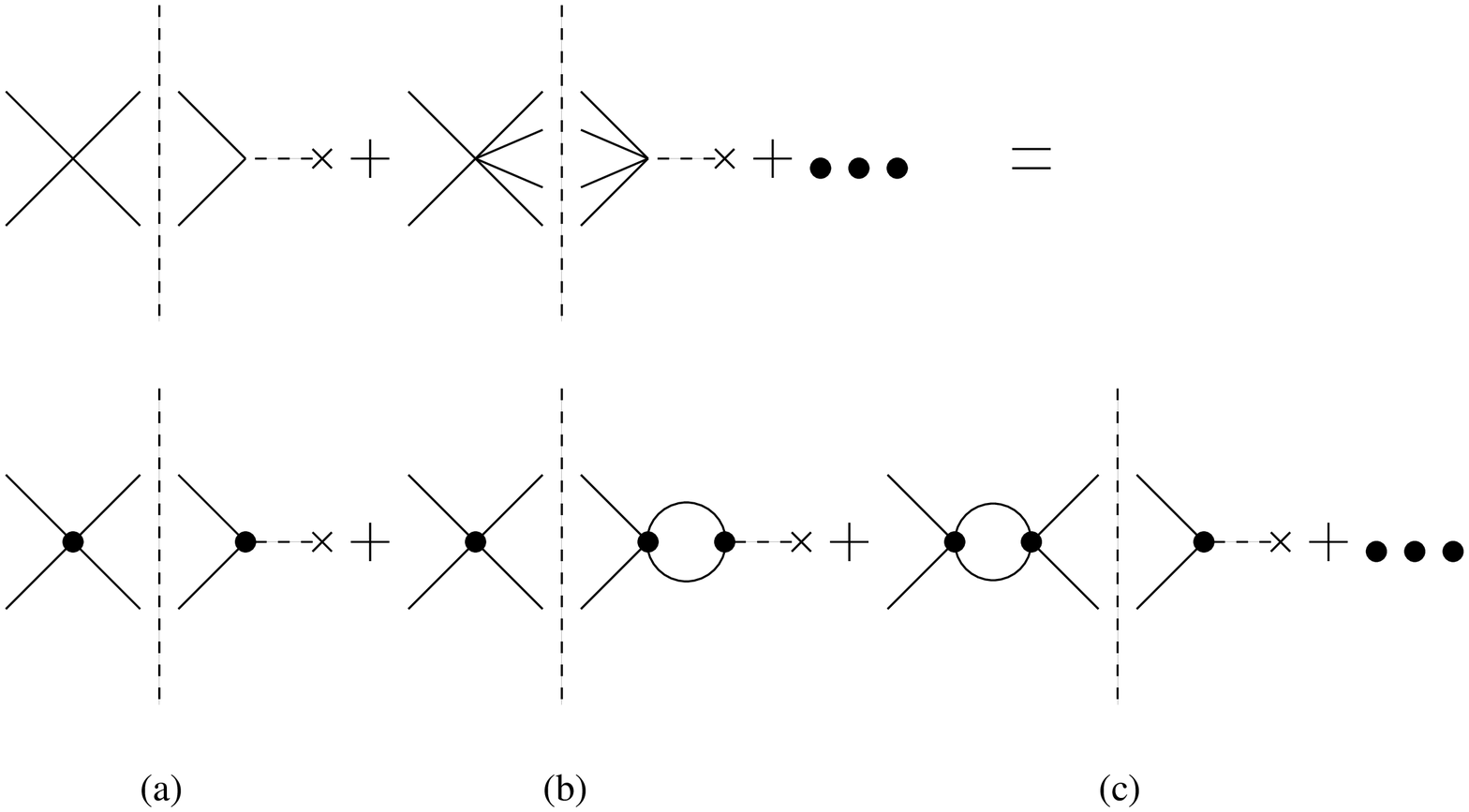}
}
\caption{\protect Imaginary part of $\Gamma_K$. The $\times$ denotes the
coupling to the scalar-isoscalar source.
\label{Zerlegung}}
\end{figure}

\medskip\noindent
The form factor
$\Gamma_K$ can  be represented by means of 
an $n$--fold subtracted dispersion relation, which, restricted to the real axis 
below respectively above the upper rim of the two-pion cut beyond threshold, reads:
\begin{equation}
\Gamma_K (s+i\epsilon)=\sum_{i=0}^{n-1} a_i s^i+\frac{s^{n}}{\pi} 
\int_{4 M_{\pi}^2}^{\infty} \frac{d s'}{s'^{n}}\frac{{\rm Im}(\Gamma_K(s'+i 
\epsilon))}{s'-s-i\epsilon}~,\quad  \epsilon \rightarrow 0^+~, 
\label{intdis}
\end{equation}
where the $a_i$ are  subtraction constants, whose number $n$ is dictated by 
the convergence behavior of $\Gamma_K$ at infinity, and the $s+i\epsilon$ 
notation indicates that, for $s>4 M_{\pi}^2\,$, we evaluate $\Gamma_K$ at the 
upper edge of the branch cut. From quark counting rules, one expects the
real (imaginary) part of $\Gamma_K$ to fall off as $1/s \, (1/s^2)$. The
central object in the dispersion relation Eq.~(\ref{intdis}) is the
absorptive part, which can expressed via
\beq
{\rm Im}~\Gamma_K\bigl(s\bigr) = \frac{i}{2}\sum_{n}\langle K^+(p_3),
\, K^+(p_1)\, {\rm in}|{\cal T}^{\dagger}|{\rm in}\,n\rangle\langle n\,{\rm in}| 
{\cal T}\,J|0\rangle ~,
\label{uni}
\eeq
where the summation extends over the complete set of intermediate states
$|n\rangle\langle n|$, i.e. 
including all sorts of multi--particle states with appropriate quantum numbers to 
satisfy the pertinent conservation laws \footnote{Note that from here
on we label the incoming momenta as $p_1$ and $p_3$ and the out--going
ones as $p_2$ and $p_4$, see also appendix~\ref{app:forder}.}. 
Furthermore, $J = \hat m (\bar u u
+\bar d d)$ is the scalar-isoscalar source (current) under consideration, and  
we have made use of the ${\cal T}$--operator, 
which is the non--trivial part of the ${\cal S}$--operator transforming a state 
$|{\rm in}\rangle$ from the Fock space of incoming states into an outgoing state 
$|{\rm out}\rangle\,$, ${\cal S}|{\rm in} \rangle= (1+i{\cal T})|{\rm in} 
\rangle=|{\rm out}\rangle\,$.
The second term on the right-hand-side of Eq.~(\ref{uni}) is  nothing but the
complex conjugate of that scalar form factor describing the coupling of the source 
$J$ to the particles of the intermediate state labeled $n\,$, while the first 
term is essentially the amplitude associated with two--kaon scattering into this
particular 
intermediate state. We have thus reexpressed the imaginary part of $\Gamma_K$ in 
terms of various form factors and scattering amplitudes. In an order by order 
analysis it follows that the lowest order imaginary part of the scalar form factor 
is of order ${\cal O}(p^4)\,$, since on the right hand side of Eq.~(\ref{uni}) there 
are two quantities of at least second chiral order. More generally speaking, 
${\rm Im}~\Gamma_K$ to any order $d$ in the energy expansion is related to 
${\rm Re}~\Gamma_K$ to  order $d-2\,$. ${\rm Im}~\Gamma_K$ to order $d$ 
is completely determined by the Lagrangian terms up to order $d-2$ via 
Eq.~(\ref{uni}). Therefore, once its imaginary part to order $d$ is known, 
we can, within the  analyticity domain, recover the order ${\cal O}(p^d)$ 
contribution of $\Gamma_K$  up to a number of subtraction constants by invoking 
the analyticity properties  of its perturbative expansion. To leading order
in the chiral expansion, we have to consider two-particle intermediate states
in Eq.~(\ref{uni}). It is well established from phenomenology that four particle 
(pion) intermediate states only play a role for energies above about 1.3~GeV and 
will thus be neglected in what follows.
For the case under consideration, the following isospin
zero states made from two equal Goldstone bosons must be considered,
\begin{eqnarray}
\pi^+\,\pi^-+\pi^-\,\pi^+-\pi^0\,\pi^0 &=& -\sqrt{3}\,|0,\,0\rangle~,\nonumber \\
K^+\,K^--K^-\,K^+-\overline{K^0}\,K^0+K^0\,\overline{K^0}&=& 2 \sqrt{2}\,|0,\,0
\rangle~,\nonumber \\  \eta\,\eta&=& |0,\,0\rangle~.
\label{mischko}
\end{eqnarray}
Performing furthermore the S-wave projection of the corresponding $KK \to
\pi\pi, KK, \eta\eta$ scattering
amplitudes (since we are dealing with a scalar source), the imaginary part
of $\Gamma_K$ at sixth chiral order can  finally be written as:
\begin{eqnarray} 
{\rm Im}~\Gamma_K^{(6)}(s) &=&-\sqrt{\frac{3}{2}} \Bigl(t^{0,(2)}_{0,K K 
\rightarrow \pi \pi}{\rm Re}~ \Gamma_{\pi}^{ (4)}  + {\rm Re}
\bigl(t^{0,(4)}_{0,K K \rightarrow \pi \pi}\bigr) \Gamma_{\pi}^{ (2)}\Bigr)
\Sigma_{\pi}\bigl(s\bigr)   \Theta\bigl(s-4 M_{\pi}^2\bigr)\nonumber\\
& &\quad +2\Bigl( t^{0,(2)}_{0,K K \rightarrow KK}{\rm Re}~\Gamma_{K}^{ (4)}
+{\rm Re}\bigl( t^{0,(4)}_{0,K K \rightarrow KK}\bigr) \Gamma_{K}^{ (2)}\Bigr)
\Sigma_{K}\bigl(s\bigr)   \Theta\bigl(s-4 M_{K}^2\bigr)\nonumber\\
& & +\frac{1}{\sqrt{2}}\Bigl( t^{0,(2)}_{0,K K \rightarrow \eta \eta}
{\rm Re}~\Gamma_{\eta}^{ (4)} + {\rm Re}\bigl(t^{0,(4)}_{0,K K \rightarrow 
\eta \eta}\bigr) \Gamma_{\eta}^{ (2)}\Bigr)\Sigma_{\eta}\bigl(s\bigr)   
\Theta\bigl(s-4 M_{\eta}^2\bigr)~,
\label{impart}
\end{eqnarray}
where we have generalized our definitions of the (non-strange) scalar form factors
according to
\beq
\langle \Phi_a(p_3),\, \Phi_b(p_1)\, {\rm out}|J|0 \rangle=\delta_{ab} 
\Gamma_a(s)~.
\label{neff}
\eeq
For $a \in \{4,5,6,7\}\,$, 
Eq.~(\ref{neff}) coincides with the earlier definition of $\Gamma_K\,$. The scalar 
form factor $\Gamma_{\pi}$ of the pion corresponds to $a \in \{1,2,3\}\,$, 
where there is no need to distinguish these three cases in the isospin symmetric 
case. The scalar form factor $\Gamma_{\eta}$ of the eta results from the choice 
$a=8\,$, when mixing of $\Phi_3$ and $\Phi_8$ is neglected.   The superscripts
$(n)\, (n = 2,4)$ refer to the chiral order, that is to tree level (2) and
one-loop accuracy (4). Furthermore, $\Sigma_a (s) = \sqrt{1-4M_a^2/s}$ and
$t^0_0$ denotes the corresponding $l=0$, $I=0$ scattering amplitudes. 
A graphic illustration of this formula in provided by fig.~\ref{Zerlegung}. 

\noindent
From this formula, the ingredients necessary for the calculation of Im~$\Gamma_K$ 
can be read off: Besides $\Gamma_K\,$, we 
need the scalar form factors of the pion and of the eta to one-loop order, 
furthermore the S-wave projections of the isospin zero amplitudes 
$T^{0}_{KK  \to \pi \pi},\,T^{0}_{KK \to KK}\,$, 
and $T^{0}_{KK \to \eta \eta}$ to ${\cal O}(p^4)$. 
All these quantities are listed in 
appendix~\ref{app:forder}. Note that for calculating the imaginary part 
of $\Gamma_K$ given in Eq.~(\ref{impart})
we need the $KK \to \pi\pi$ scattering amplitude in the unphysical region
$s \in [4M_\pi^2 , 4M_K^2]$. The amplitude can be reconstructed in this
regime by means of an Omn\`es representation, as detailed in \cite{ABM}.
We refer to that paper for all details and simply apply the same procedure.

\medskip\noindent
In what follows, we chose to work with a triple
subtracted dispersion relation for the scalar kaon form factor $\Gamma_K (s)$. 
Therefore, the normalization, the radius and the curvature terms appear in the 
polynomial part of the dispersive representation, which allows for the 
most transparent way of fixing the various subtraction constants (LECs).
We   thus have
\begin{equation}
\Gamma_K^{(4+6)} \bigl(s+i\epsilon\bigr) = P (s) 
+\frac{s^{3}}{\pi} \int_{4 M_{\pi}^2}^{\infty} \frac{d s'}{s'^{3}}
\frac{{\rm Im}~\Gamma_K^{(4+6)}(s'+i \epsilon)}{s'-s-i\epsilon}~,
\label{scons}
\end{equation}
with the polynomial
\beq
P(s) = P_4(s) + P_6(s) =
\frac{M_{\pi}^2}{\Lambda_\chi^2} \, \biggl(
\left(d_1 M_\pi^2 + \frac{d_2 M_{\pi}^4}{\Lambda_\chi^2}\right) 
+ \left(f_1 + \frac{f_2 M_{\pi}^2}{\Lambda_\chi^2}\right) s+ 
\frac{g}{\Lambda_\chi^2} \frac{s^2}{2}\biggr)~.
\eeq
Here, the dimensionless numbers $d_1, f_1 \, (d_2,f_2,g)$ are combinations of 
dimension four (six) LECs. In what follows, we will fix $d_1$ and $f_1$ from
the normalization and radius at one-loop accuracy and set $d_2 = f_2 =0$ for
our central results. We will also vary the latter two within reasonable
bounds, $\Delta d_2 = \Delta f_2 = \pm 1/(16\pi)^2$. The coupling $g$ can be
determined from the requirement that the normalized scalar form factor 
$\Gamma_K / M_\pi^2$ stays finite in the chiral limit (cl). 
Setting $m_u = m_d = m_s = 0$, 
we find the following representation of the sixth order contribution
to this quantity:
\begin{eqnarray}
\frac{ \Gamma_K^{(6),{\rm cl}}\bigl(s\bigr)}{ M_{\pi}^2}
&=&  \frac{1}{(4\pi F)^4} \, s^2 \, \biggl(\bigl(4\pi\bigr)^2
\bigl(\frac{2632}{45}L_1^r+\frac{3082}{135}L_2^r+\frac{8773}{405}L_3^r
+\frac{70}{3}L_4^r+\frac{1012}{135}L_5^r  
-\frac{428}{45}L_6^r
-\frac{85}{27}L_8^r\bigr)\nonumber \\ & &
+\bigl(4\pi\bigr)^2\bigl(\frac{68}{3}L_1^r+\frac{88}{9}
 L_2^r+\frac{232}{27} L_3^r+12 L_4^r+\frac{7}{2}L_5^r\bigr)
\, \Bigl(\ln\bigl(\frac{M^2}{\lambda^2}\bigr)
+\ln\bigl(\frac{\lambda^2}{-s}\bigr)\Bigr) 
\nonumber\\ & &
+\frac{g}{2}+\gamma+\frac{17761}{12960}
-\frac{97871}{25920}\ln\bigl(\frac{M^2}{\lambda^2}\bigr)-\frac{325}{192}\ln^2
\bigl(\frac{M^2}{\lambda^2}\bigr)+\frac{661}{192}\ln\bigl(\frac{\lambda^2}{-s}\bigr)
+\frac{325}{192}\ln^2\bigl(\frac{\lambda^2}{-s}\bigr)\biggr) ~,
\end{eqnarray}
where the constant $2\gamma/(4\pi F)^4$ is the part of the second derivative
of the absorptive part of the dispersive representation of $\Gamma_K$
with respect to $s$ generated by the terms in the S-wave projected 
scattering amplitudes which we could only represent in an integral
form, see appendix~\ref{app:forder}.
Also, the arguments of the logarithms $\sim M^2$ have been made dimensionless
by the square of the renormalization scale $\lambda$.
The requirement that $  \Gamma_K$ stays finite in the chiral limit implies 
that the chiral logarithms are compensated by corresponding terms in $g\,$:
\beq
g = g^r-\bigl(4\pi\bigr)^2\biggl(\frac{136}{3}L_1^r+\frac{176}{9} L_2^r
+\frac{464}{27} L_3^r+24 L_4^r+7L_5^r\biggr)\ln\biggl(\frac{M^2}{\lambda^2}\biggr)
+\frac{97871}{12960}\ln\biggl(\frac{M^2}{\lambda^2}\biggr)
+\frac{325}{96}\ln^2\biggl(\frac{M^2}{\lambda^2}\biggr)~,
\label{zzz}
\eeq
where parts of $\gamma$ have been reshuffled to the finite constant $g^r$.  
$g$ is thus found to contain chiral logarithms and squared chiral logarithms 
together with a finite part.  This structure reflects the singularity structure 
of $g$ before renormalization: In dimensional regularization, 
$g$ absorbs poles of first and of second order in $1/(d-4)\,$, with the related 
chiral logarithms restoring independence on the renormalization scale. 
The scale-dependence of  $g$ is given by
\begin{equation}
\frac{\partial g}{\partial \lambda}=\frac{\partial g^r}{\partial \lambda}
+ \frac{1}{\lambda}\Biggl((4\pi\bigr)^2\left(\frac{272}{3}L_1^r
+\frac{352}{9} L_2^r+\frac{928}{27} L_3^r+48 L_4^r+14L_5^r\right)
-\frac{97871}{6480}\Biggr)~,
\end{equation}
i.e.\ the derivative of the double chiral logarithm is canceled by the contributions 
of the $L_i\,$. Similarly, the logarithmic scale dependence of the $L_i$
balances the scale dependence of $g^r$, $\lambda  \partial g^r/\partial  \lambda$, 
when requiring $\partial g/\partial  \lambda=0\,$. Eq.~(\ref{zzz}) allows to estimate
the coupling $g$. 
Neglecting $g^r$, we evaluate $g$ for $\lambda=M_{\rho}\,$. Identifying the meson mass
$M$ with the pion (eta) mass, we find $g=6.7 \, (-5)$. This ambiguity is to be 
contrasted with the two flavor case, where only the pion mass can appear, and thus
the corresponding constant can be fixed unambiguosly \cite{GaMe}. Since the
chiral pion loops are longer ranged than kaon or eta loops
(and are thus more important), it is however reasonable
to set $M = M_\pi$ also in the SU(3) case as will be done in what follows.
\medskip\noindent

\begin{figure}[htb]
\epsfysize=2.5in
\epsffile{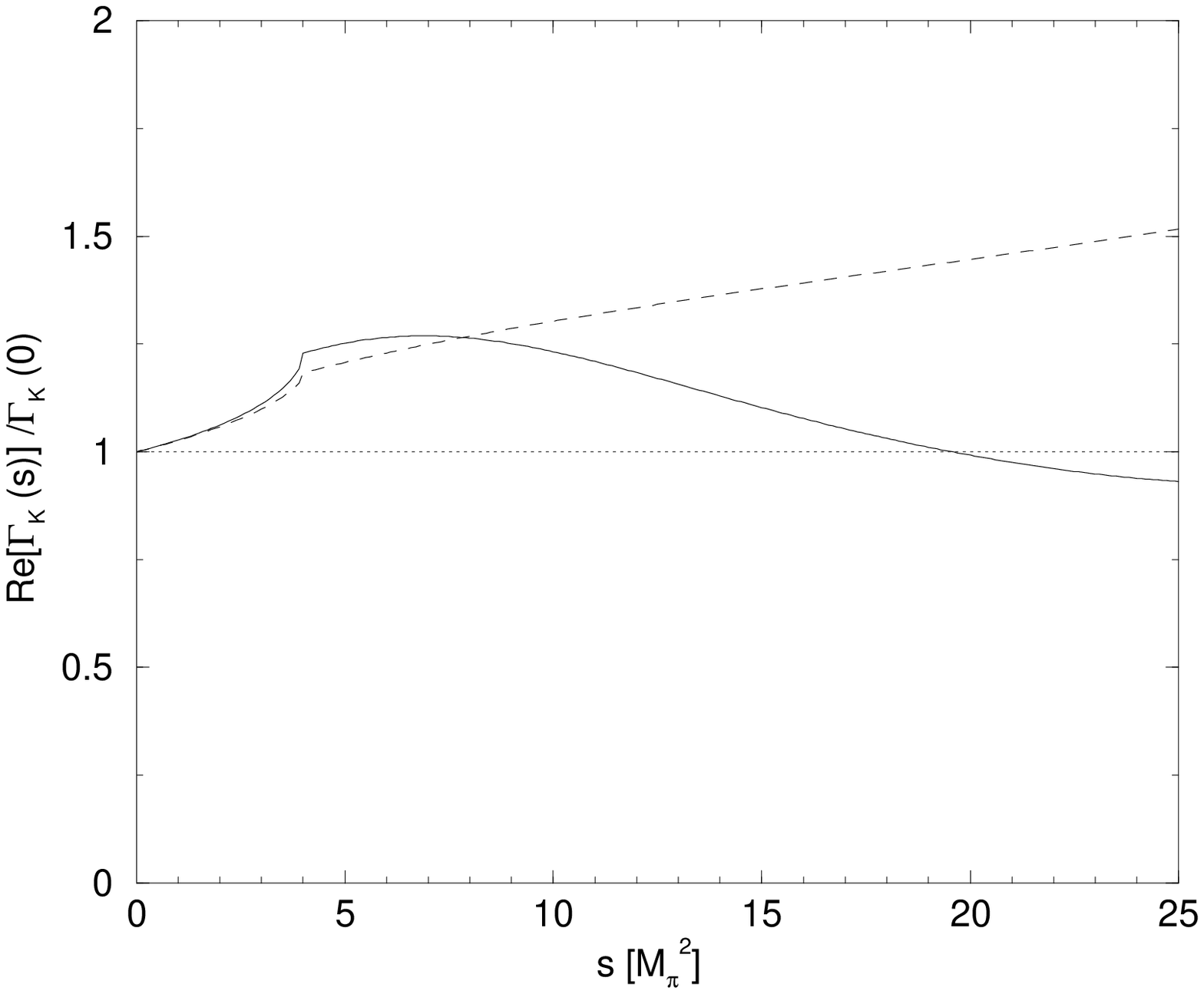}

\vspace{-2.5in}

\hfill
\epsfysize=2.5in
\epsffile{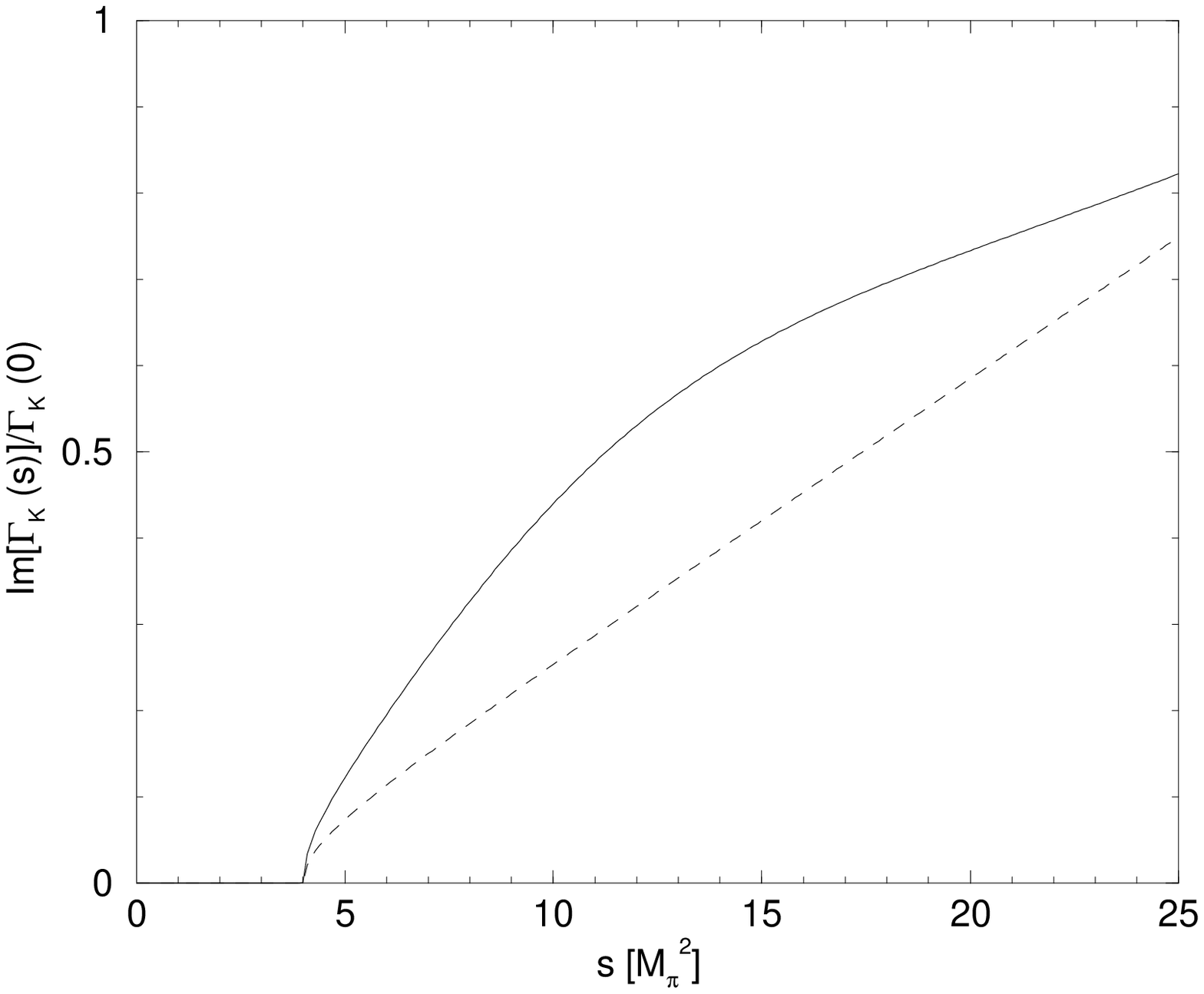}
\caption{\protect Normalized scalar form factor $\Gamma_K (s)/\Gamma_K (0)$. 
Left (right) panel: Real (imaginary) part. The dotted, dashed and solid lines
represent the tree, one-loop and two-loop result, in order.
\label{fig:sff}}
\end{figure}
\noindent
The real and imaginary parts of the normalized (non-strange) scalar kaon form factor
$\Gamma_K$
are shown in fig.~\ref{fig:sff}. Consider first the real part. The overall correction
to the tree level result at the two-pion threshold $s = 4M_\pi^2$ amounts to 23\%, with
only 5\% due to two-loop effects. For larger energies, the two-loop result turns over
while the one-loop curve keeps on rising. This is similar to the case of the pion
scalar form factor \cite{GaMe}. The turn-over of the two-loop curve is due to the
one-loop phase passing through 90$^\circ$ in the region of the scalar resonances.
If one varies the constants $d_2$ and $f_2$ as described before, the real part is
only mildly affected. Its decrease for $s \geq 10\,M_\pi^2$ is much steeper if we chose
to set $M= M_\eta$ in the determination of the constant $g$. However, at $s=4 M_\pi^2$,
this different choice of $g$ only reduces the two-loop correction to about 3\%. 
The imaginary part is only non-zero at one-loop order and the corrections 
from the two-loop graphs are more sizeable in the threshold region
as shown in the right panel of fig.~\ref{fig:sff}. This is very similar
to the case of the scalar pion form factor studied in \cite{GaMe}.
Note also that the final state interactions are weaker in the
pion--kaon system than in the pion case (as signalled e.g. by the
mass of the dynamically generated light scalar mesons in the two
channels, see e.g. \cite{JOP}). 
We stress again the difference to the case of pion--nucleon scattering.
There, the momentum dependence is much stronger (for the reasons discussed
in the previous section), despite the apparent similarity to the case of
pion--kaon scattering considered in this work.
 
\section{Summary}
\def\theequation{\arabic{section}.\arabic{equation}}
\setcounter{equation}{0}
\label{sec:summ}
To summarize, we have considered aspects related to scalar form factors
and pion-kaon scattering in chiral perturbation theory. More precisely,
the pertinent results of this investigation can be summarized as follows:
\begin{itemize}
\item[(1)]We have analyzed the low--energy theorem Eq.~(\ref{CDK}) 
for pion--kaon scattering. The remainder at the Cheng-Dashen point
turns out to be much smaller than expected from naive dimensional analysis
in three flavor chiral perturbation theory. In particular, setting the meson
decay constant $F=F_\pi$, the remainder is comparable to the one in
pion--pion scattering \cite{GS}.
\item[(2)]We have shown that the result for the remainder can be understood
in terms of approximate scale relations by representing  the one--loop
corrections to the scalar kaon form factor and the $\pi K$ scattering amplitude
in terms of chiral logarithms with appropriate scales $\Lambda_i$ and 
$\tilde{\Lambda}_i$, $i=1,2$. These
scales come out to be close to the eta mass, thus suppressing the potentially
large chiral logarithms multiplying the kaon mass squared. 
\item[(3)]We have repeated the analysis in the heavy kaon framework, in which 
the kaons are treated as matter fields. Matching conditions allow to fix
the new low--energy constants from the ones based on the standard chiral
expansion with light kaons. We have performed this matching procedure for
the scalar kaon form factor and the $\pi K$ scattering amplitude. This
analysis confirms the finding in the standard approach. Since a heavy kaon
can not decay, no ambiguity arises as to the choice of the meson decay constant.
\item[(4)]Since the pion scalar form factor is subject to large two--loop
corrections already close to threshold, we have calculated these also for
the scalar kaon form factor using a dispersive representation \cite{GaMe}.
The pertinent subtraction constants are fixed from the one--loop representation
of this form factor and the condition that it is well defined in the chiral
limit. The resulting two--loop corrections for the real part are fairly small at 
low energies, while they are more pronounced for the imaginary part. We have
also discussed the dependence of the scalar kaon radius on the OZI--violating
low--energy constant $L_4$.
\end{itemize}

\bigskip  

\section*{Acknowledgments}
We are grateful to J\"urg Gasser for some valuable comments and
encouragement. We also thank Paul B\"uttiker for tuition on
subthreshold scattering amplitudes. 

\bigskip  

\appendix
\def\theequation{\Alph{section}.\arabic{equation}}
\setcounter{equation}{0}
\section{Low-energy kaon relations}
\label{app:A}
In very much the same manner in which we have analyzed the isospin 
even $\pi K$ amplitude in the regime of low pion momenta, we can 
also consider the limit of vanishing kaon momenta. To be specific, we 
consider again the reaction $K^+(p_1) + \pi^+(p_3) 
\rightarrow K^+(p_2) + \pi^+(p_4)$, but now in  the soft-kaon limit of vanishing
kaon four-momenta. The resulting low-energy theorem for the on--shell 
amplitude reads
\begin{equation}
T^{+}_{\pi K\to \pi K}(\nu=0,\, t=2M_{K}^2) \equiv
\tilde{A}_{\pi K}^{\rm CD} = \frac{1}{F^2}\tilde{
 \Gamma}_{\pi^+}(2 M_{K}^2)+{\cal O}(M_{K}^4)~,
\label{KLe}
\end{equation}
where the strange form factor of the pion, 
$\tilde{ \Gamma}_{\pi}$, is defined as follows:
\begin{equation}
\tilde{\Gamma}_{\pi}(t)=\langle \pi^+(p_4)|\frac{1}{2}\bigl(
\hat{m}+m_s\bigr) \bigl(\bar{u}u+\bar{s}s\bigr)|\pi^+(p_3) \rangle~,
\label{strangeff}
\end{equation}
with $t=(p_3-p_4)^2\,$.
The analogue of the CD--point is now the kinematic configuration 
$ \nu=0,\,t=2 M_K^2$ far off the physical region of elastic scattering. 
That makes the experimental determination of $\tilde{ \Gamma}_{\pi}$ 
more involved than in the case of $\Gamma_K$ since data have to be 
extrapolated further beyond the physical domain.
The one-loop calculation of $\tilde{ \Gamma}_{\pi}$ is 
straightforward and leads to:
\begin{eqnarray} 
\tilde{\Gamma}_{\pi^+}(t)
&=&\frac{M_{K}^2}{2}+\frac{M_{K}^2}{2F^2}\Bigl[L_4^r
\bigl(-32M_{\pi}^2+16 t\bigr)+L_{5}^r
\bigl(-16M_{\pi}^2+8 M_{K}^2+4t\bigr)+L_6^r 64 M_{\pi}^2
+L_8^r\bigl(32M_{\pi}^2-16 M_{K}^2\bigr) 
\nonumber \\& & 
+\frac{1}{2}M_{\pi}^2\mu_{\pi}
+\bigl(\frac{1}{6}M_{\pi}^2-\frac{2}{3} M_K^2\bigr)\mu_{\eta}
+\bigl(\frac{1}{2}M_{\pi}^2-t\bigr) J^r_{\pi \pi}\bigl(t\bigr)
-\frac{3}{4}t J^r_{KK}\bigl(t\bigr) - \frac{5 }{18}M_{\pi}^2 
J^r_{\eta \eta}\bigl(t\bigr)\Bigr]+{\cal O}(p^6)~.
\end{eqnarray}
Note that the OZI violating coupling $L_4$ contributes significantly to the
form factor for four-momenta of the order $t \simeq M_K^2$, but is suppressed in
the low-energy region much as the other OZI violating LEC, $L_6$,
as it only appears with a prefactor $\sim M_\pi^2$.
The first moment of the low momentum expansion of this form factor is
given in terms of the pertinent strange pion radius,
\beq\label{rpi}
\langle \tilde{r}_S^2 \rangle_\pi
        = \biggl\{ \begin{tabular}{ll}
        $(0.41 \pm 0.22)$~{\rm fm}$^2$ & $\quad$ {\rm set}~1~, \\
        $(0.49 \pm 0.02)$~{\rm fm}$^2$ & $\quad$ {\rm set}~2~.  
        \end{tabular} 
\eeq
whose central value is smaller than the corresponding non--strange radius
of about 0.6~fm$^2$. This pattern is to be expected since the strange
quark is more massive than the light quarks and thus leads to smaller
scales in coordinate space. Concerning the theoretical uncertainty,
the same remarks as after Eq.~(\ref{rK}) apply here.
In analogy with our previous considerations, we examine the validity of the 
low--energy theorem Eq.~(\ref{CDK}) in terms of the remainder $\tilde{\Delta}_{\pi K}\,$:
\begin{equation}
F^2 \,  \tilde{A}_{\pi K}^{\rm CD} = 
\tilde{ \Gamma}_{\pi} (2 M_{K}^2) + \tilde{\Delta}_{\pi K}~.
\end{equation}
Since we are now working above the two pion threshold $t=4 M_{\pi}^2\,$, 
we will generally have to deal with imaginary contributions.
As it turns out, these cancel exactly in the 
low--momentum kaon relation, such that $\tilde{\Delta}_{\pi K}$ is real. 
The numerical results for the amplitude and for $\tilde{\Gamma}_{\pi}$ 
at $\nu=0,\,t=2 M_K^2$ are displayed in units of $M_K^2$ in 
table~\ref{tab:rem1}, together with the remainder in units of $M_K^2$ and 
its relative size $R\,$, defined as the ratio of $|\tilde{\Delta}_{\pi K}|$ and 
the complex modulus of the amplitude at $\nu=0,\,t=2 M_K^2\,$. 
$R$ is given in $\%\,$.
\begin{table}[htb]
\begin{center}
\begin{tabular}{|l|c|c|c|c|c|}
    \hline
$F^2$       & $L_i$ set 
            &  $F^2 \tilde{A}_{\pi K}^{\rm CD} $ [$M_K^2$]
            & $\tilde{\Gamma}_{\pi}(2 M_{K}^2)$ [$M_K^2$]
&$\tilde{\Delta}_{\pi K} $[$M_K^2$] & $R$ [\%] \\
    \hline
$F_\pi^2$   & 1 & $0.779+0.510i $   &  $0.825+0.510i $ & $-0.046$   
&  $4.9$     \\
$F_\pi F_K$ & 1 & $0.886+0.510i $   &  $0.825+0.510i $ & $0.061$   
&  $6.0$     \\
$F_K^2$     & 1 & $0.993+0.510i $   &  $0.825+0.510i $ & $0.168$   
&  $15.1$     \\
$F_\pi^2$   & 2 & $0.813+0.510i$    &  $0.917+0.510i $ & $-0.104$   
&  $10.8$     \\
$F_\pi F_K$ & 2 & $0.894+0.510i $   &  $0.917+0.510i $ & $-0.021$   
&  $2.2$     \\
$F_K^2$     & 2 & $0.975+0.510i $   &  $0.917+0.510i $ & $0.058$   
&  $5.3$     \\
\hline\hline
\end{tabular}
\centerline{\parbox{12cm}
{\caption{Low--momentum kaon theorem. Size of the remainder 
$\tilde{\Delta}_{\pi K}$ for various choices of the
meson decay constants and the low--energy constants 
$L_i^r (M_\rho)$.
         \label{tab:rem1}}}}
\end{center} 
\end{table}
\noindent
The situation is somewhat different compared to the previous case.
For the LECs from set~1, the best agreement is given for the normalization 
$1/F_{\pi}^2$, where the relative deviation amounts to about $5$ \%. 
The situation is different when working with the LECs from set~2, where the 
decay constant combination $1/F_{\pi} F_K$ accounts for the smallest 
remainder of about $2$ \% relative size. Both of these values for 
$R$ are larger than the corresponding values in the low-energy pion case,
as it is expected due to the larger kaon mass. Also, the dependence on the
choice of decay constant is less pronounced. We note, however, that the 
deviations from the LET are fairly small for all choices of $F^2$ and
are well below the typical SU(3) corrections $M_K^2 /\Lambda^2_\chi \simeq
0.2$.

\def\theequation{\Alph{section}.\arabic{equation}}
\setcounter{equation}{0}
\section{Basics of heavy kaon chiral perturbation theory}
\label{app:heavyK}
The basic concepts of heavy kaon CHPT (HKCHPT) are adopted
from heavy baryon CHPT, as introduced in Ref.~\cite{JM} to include baryon fields 
into the framework of chiral peturbation theory.
Since baryon masses are comparable to the chiral symmetry 
breaking scale and are non--vanishing in the chiral limit, they cannot 
be considered light. As a  consequence 
hard momenta enter into the theory and the standard power counting scheme 
breaks down. This is because arbitrarily complicated diagrams no longer 
yield contributions of a fixed chiral order, but contributions 
of any lower order are now possible if only a sufficient number of momenta 
is provided by derivatives acting on the heavy fields. Heavy baryon CHPT
therefore treats baryons essentially as static in an extreme non--relativistic 
framework with small residual (that is, soft) momenta.
In the standard approach, see e.g. Ref.~\cite{bkkm}, 
a baryon field $B(x)$ is rewritten in the form
\begin{equation}
B(x)=e^{-i m v \cdot x} b(x)~,\label{HK}
\end{equation}
where $m$ is the baryon mass and $v_{\mu}$ a four--velocity obeying 
$v^2=1\,$. The field $b$  has only small residual momentum 
which can be treated on an equal footing with the other generically 
small momenta and masses, $p$. Using Eq.~(\ref{HK}) for the heavy
fields, one can perform an expansion in powers of $1/m\,$. 
The result is a Lagrangian which generally breaks Lorentz invariance 
and gives rise to a modified propagator with 
additional new vertices suppressed by powers 
of $1/m\,$. These can be included in a  power counting scheme 
where contributions are organized in terms of both powers of 
$p/\Lambda_{\chi}$ and of $p/m\,$. Since $\Lambda_{\chi}$ and 
$m$ are of the same order of magnitude,  it is not necessary to 
differentiate between these various types of contributions.

\medskip\noindent
In a similar way, consider now the kaon to be a heavy particle on the pionic mass 
scale, and apply a similar  scheme to pion-kaon reactions. As before, the
heavy mass scale (now the kaon mass) has to be eliminated to allow for
a consistent power counting (if one uses conventional dimensional regularization).
We will closely follow the approach presented 
in Ref.~\cite{Roessl}, correcting for a number of apparent misprints, and adding
some new results.
It is clear that a theory which treats pions as light, relativistic 
particles and kaons as heavy, non--relativistic ones cannot respect 
SU(3) symmetry.  The pertinent symmetry group 
will therefore be SU(2)$_V \times $SU(2)$_A\,$. We therefore have to
choose different representations for the pion and the kaon fields 
and construct the most general Lagrangian compatible with the symmetries 
of QCD, where again chiral symmetry plays a prominent role. Lorentz 
invariance will require special attention. Since the kaon now plays the
role of any matter field in a theory with non-linearly realized chiral symmetry,
it is natural to apply the CCWZ formalism \cite{CCWZ}. For doing that,
we combine the kaon fields into a representation as isospin doublets :
\begin{equation}
K= \left(\begin{array}{c}K^+\\K^0\end{array}\right)~, \quad
\tilde{K}= \left(\begin{array}{c}-\overline{K^0}\\K^-\end{array}\right)~.
\end{equation}
In what follows, we will write $K$ as a generic symbol for any of these doublets
(and call them kaon fields).
The advantage of this representation is that the compensator field $h$ provides 
a natural way to define the action of SU(2)$_R \times $SU(2)$_L$ on the kaon fields:
\begin{equation}
K(x) \rightarrow h(R,\,L,\,u(x))\, K(x)
\end{equation}
for $ R/L \in~$SU(2)$_{R/L}\,$ and $u^2(x) = U(x)$ parameterizes the
Goldstone boson fields. Note that
for pure vector transformations $R=L\,$, $h$ simplifies to $h=R=L\,$, 
so that SU(2)$_V$ is represented fundamentally on $K$.
A striking difference in this treatment of the Goldstone bosons 
is the non--linear representation for the pion as against the 
linear one for the kaon degrees of freedom. While the former allows for 
dealing with multi--pion couplings by an expansion in powers of the relevant 
fields, from the latter it immediately follows that the theory will fall 
into separate sectors marked by the occurrence of a fixed number of kaons, 
therefore the effective Lagrangian ${\cal L}_{\rm HKCHPT}$ can be written
as a string of terms:
\begin{equation}
{\cal L}_{\rm HKCHPT}={\cal L}_{ \pi}+ {\cal L}_{\pi KK}+ 
{\cal L}_{\pi KK K K}+\ldots~.
\end{equation}
While the first term describes purely pionic processes, the second one is 
bilinear in the kaon field, and the third one is quadrilinear, and so on.
In this paper, we only consider processes with one in-coming and one out-going 
$K\,$, i.e. only the first two terms in this series will be of relevance.
In this framework closed kaon loops, i.e. loops formed by internal $K$ lines
only, are prohibited, and their effects enter implicitly, absorbed into 
the coupling constants. Kaon propagators do, however, show up in loops composed 
of both $\pi$ and $K$ internal lines and thus the large mass scale $M_K$ destroys
the power counting. To remedy this, one could proceed as outlined above, 
i.e.\ go over to the extreme non--relativistic limit via a field 
transformation analogous to Eq.~(\ref{HK}), 
\beq
K(x)=e^{-i M_K v \cdot x} k(x)~,
\label{HKK}
\eeq 
and give up Lorentz invariance right on the Lagrangian level. Diagrams 
are then calculated in a non--relativistic framework, and Lorentz invariance 
is invoked at a later stage to determine a number of relations among the 
coupling constants. Indeed Roessl \cite{Roessl} lists the most general Lagrangian in the 
fields $u$ and $k$ up to fourth order in  small momenta, compatible 
with the symmetries of QCD except for Lorentz invariance. However, to 
perform calculations a different approach is proposed (such a modified
scheme has also been applied in calculations of heavy baryon CHPT): One determines the
manifest Lorentz invariant Lagrangian in terms of the fields $u$ and $K$ 
which generates those non--relativistic ones via the relation Eq.~(\ref{HKK}). 
Calculations are then relativistically invariant at any stage up to the 
evaluation of loop integrals. Only then the heavy particle expansion in 
$1/M_K$ is performed in those integrands containing heavy propagators, 
i.e.\ integrals of the type (as a typical example, consider a loop function
with one pion and one (heavy) kaon propagator):
\begin{equation}
J_{\pi K}\bigl((p_1-p_2)^2\bigr)=\int \frac{d^d k}{(2 \pi)^d} 
\frac{i}{k^2-M_{\pi}^2}\frac{1}{(p_1-p_2-k)^2-M_K^2} ~. 
\label{sosehr}
\end{equation}
Since this expression is Lorentz invariant, one is free to work in a
frame where the incoming kaon momentum, say $p_2\,$, is of the form
$p_2 = M_K \, v= (M_K,0,0,0)\,$.
Plugging this into the integral and expanding the integrand in powers of 
$ 1/M_K$ then yields
\begin{equation}
J_{\pi K}\bigl((p_1-p_2)^2\bigr)=\int \frac{d^d k}{(2 \pi)^d} 
\frac{i}{k^2-M_{\pi}^2} \Bigl(-\frac{1}{2 v \cdot (p_1-k)} \frac{1}{M_K}
-\frac{(p_1-k)^2}{4[v \cdot (p_1-k)]^2} \frac{1}{M_K^2} +\ldots \Bigr)~,
\label{ser}
\end{equation}
where the ellipsis indicates higher powers in $1/M_K\,$. So one ends up with 
a series of  terms organized as an expansion in $p/M_K\,$, where $p$ is a
generic small CHPT scale. On a diagrammatic level these can be 
represented by absorbing the first contribution into a modified propagator, 
and treating the remaining ones as additional vertices of proper order. We 
can then arrange any perturbative expansion derived from 
${\cal L}_{\rm HKCHPT}$ as a dual expansion in powers of both 
$p/\Lambda_{\chi}$ and $p/M_K\,$. The pertinent  power counting rules can be easily
derived.  Consider the amplitude ${\cal A}$ of
an arbitrary graph consisting of $V_n^{\pi \pi}$ pionic vertices of order 
$n\,$, $V_m^{\pi K}$ pion--kaon vertices of order $m\,$, $E^{\pi }$ external 
pion legs, $E^{K }$ external kaon lines, $I^{\pi}$ internal pion lines, 
$I^{K}$ internal kaon lines, and $L$ loops. The chiral dimension $\nu$ 
assigned to such a diagram is (that is, ${\cal A} \sim p^\nu$)
\begin{equation}
\nu=\sum_n V_n^{\pi \pi}(n-2)+\sum_m V_m^{\pi K}(m-1)+2L+1~,
\label{fpower}
\end{equation}
where we have used the topological identities $I^K=\sum_m V_m^{\pi K}-1$
and $I^{\pi}+I^K=L+\sum_n V_n^{\pi \pi}+\sum_m V_m^{\pi K}-1\,$.
From this equation one readily deduces that in contrast to standard
SU(3) CHPT, diagrams with odd chiral dimensions are allowed in HKCHPT.
The advantage of this 
scheme over SU(3) CHPT lies in its improved convergence properties. 
For energies of the order of the pion mass, the HKCHPT expansion 
parameter is given by $M_{\pi}/M_K \approx 0.28\,$, thus a diagram of 
order $d+2$ is suppressed relative to an order $d$ contribution by at 
least a factor $M_{\pi}^2/M_K^2 \approx 0.08\,$. This is substantially 
more favorable than the corresponding minimal suppressing factor 
$M_K^2/\Lambda_{\chi}^2\approx 0.2$ in $SU(3)$ CHPT. 
However desirable this feature may be, it is achieved at the 
price of a larger number of unknown LECs in the Lagrangian.

\medskip\noindent
The difference between the heavy kaon and the standard chiral expansion
can be clearly seen in case of the pion and the kaon masses. The pion
mass takes the canonical form,
\beq
M_\pi^2 = M^2_0 \left( 1 + 2\frac{M_0^2}{F^2}l_3^r + \frac{M_0^2}{32\pi^2F^2}
\ln \frac{M_0^2}{\lambda^2} \right)~,
\eeq
with $M_0^2 = 2\hat m B_0$ the leading term in the quark mass expansion of the
pion mass. Of course, at next-to-leading order, the SU(2) LEC $l_3^r$ appears
\cite{GLann}.
The kaon mass appears quadratically in the heavy kaon Lagrangian
and takes the form
\beq\label{MKexp}
M_K^2 = M^2 + M^{(2)} M_0^2 + M^{(4)} M_0^4~,
\eeq
where $M^2$ is the quark mass independent contribution and the explicit form of the
coefficients $M^{(2,4)}$ is given in \cite{Roessl}. Note that the first two terms
in Eq.(\ref{MKexp}) are not renormalized and thus are finite. For the matching
with the relativistic formulation one has to expand $M^2$, $M^{(2)}$ and $M^{(4)}$
in powers of $\bar{M}_K^2 = m_s B_0$. Similarly, since in the heavy kaon theory
one only has pion loops, loop functions like e.g. $J_{KK}^r$ and $ J_{\eta \eta}^r$
must be expanded in inverse powers of $\bar{M}_K^2$,
\begin{eqnarray} 
J_{KK}^r(t) &=& \frac{1}{(4\pi)^2}\biggl(1+ \ln \Bigl(  \frac{\bar{M}_K^2}{\lambda^2}
 \Bigr)+\frac{M_{\pi}^2}{2 \bar{M}_K^2}-\frac{t}{6 \bar{M}_K^2}\biggr) 
+{\cal O}\biggl(\frac{p^4}{ \bar{M}_K^4}\biggr)~,\nonumber \\ 
 J_{\eta \eta}^r(t) &=& \frac{1}{(4\pi)^2}\biggl(1+ \ln \Bigl(\frac{4}{3}  
\frac{\bar{M}_K^2}{\lambda^2} \Bigr)+\frac{M_{\pi}^2}{4\bar{M}_K^2}- 
\frac{t}{8\bar{M}_K^2} \biggr) +{\cal O}\biggl(\frac{p^4}{ \bar{M}_K^4}\biggr)~. 
\label{HKloops}
\end{eqnarray}
For more details on the heavy kaon approach, we refer to \cite{Roessl,Oul}.

\def\theequation{\Alph{section}.\arabic{equation}}
\setcounter{equation}{0}
\section{Heavy kaon CHPT Lagrangian}
\label{app:heavyKL}

First, we give the basic building blocks of the heavy kaon
Lagrangian and the associated transformation properties 
under chiral $R/L \in SU(2)_{R/L}\ $:
\beqa
\begin{array}{rclrcl}
U &=& u^2~,& U & \to& R U L^{\dagger}~,\nonumber\\
D_{\mu}K &=& \partial_{\mu}K+\Gamma_{\mu}K~, &
D_{\mu}U & \to & R D_{\mu}U L^{\dagger}~,\nonumber\\
D_{\mu \nu} U &=& \bigl( D_{\mu}D_{\nu}+D_{\nu}D_{\mu}\bigr) U~, &
D_{\mu\nu}U & \to & R D_{\mu\nu}U L^{\dagger}~,\nonumber\\
D_{\mu \nu} K &=& \bigl( D_{\mu}D_{\nu}+D_{\nu}D_{\mu}\bigr) K~, &
D_{\mu \nu}K &\rightarrow& h D_{\mu \nu}K~,\nonumber\\ 
\Delta_{\mu} &=& \frac{1}{2} u^{\dagger} D_{\mu}U u^{\dagger}~,&
\Delta_{\mu} & \rightarrow& h \Delta_{\mu} h^{\dagger}~,\nonumber \\
\Delta_{\mu \nu} &=& \frac{1}{2} \bigl( D_{\mu}\Delta_{\nu}
+ D_{\nu}\Delta_{\mu} \bigr)~, &
\Delta_{\mu \nu} & \rightarrow& h \Delta_{\mu\nu} h^{\dagger}~,\nonumber \\
 \chi_{\pm} &=& u^{\dagger} \chi u^{\dagger}\pm u 
\chi^{\dagger} u~, & \chi_{\pm} & \rightarrow& h \chi_{\pm} h^{\dagger}~,
\end{array}
\eeqa
with the chiral connection
\beq
\Gamma_{\mu} = \frac{1}{2} \Bigl(u^{\dagger} \bigl(\partial_{\mu}
-i r_{\mu} \bigr) u+u \bigl(\partial_{\mu}-i l_{\mu} \bigr) u^{\dagger} 
\Bigr)~,
\eeq
where $r_{\mu},\, l_{\mu}$ are  external right/left-handed currents.
As mentioned before, the general form of the Lagrangian up to order 
${\cal O}(p^4)$ is 
\begin{equation}
{\cal L}_{HK \chi PT}={\cal L}_{ \pi}^{(2)}+{\cal L}_{ \pi}^{(4)}+ 
{\cal L}_{\pi KK}^{(1)}+ {\cal L}_{\pi KK}^{(2)} + 
{\cal L}_{\pi KK}^{(3)}+ {\cal L}_{\pi KK}^{(4)}~,
\label{hklag}
\end{equation}
where the purely pionic sector is chosen such that
it coincides with the standard two flavor CHPT Lagrangian.
Concerning the $\pi$--$K$ interaction Lagrangian, it is clear 
from the discussion of the difficulties related to the power counting 
procedure in HKCHPT, see appendix~\ref{app:heavyK}, that there is not a 
one-to-one correspondence 
of the dimensions assigned to a Lagrangian term in the relativistic 
and the non--relativistic framework. This means that a given Lorentz invariant 
term can, via Eq.~(\ref{HKK}), give rise to contributions of different 
powers in the non--relativistic formulation. In Roessl's Lagrangian, 
the terms are labeled according to the leading non--relativistic 
contributions they lead to. The LECs are denoted $A_i,\,B_i,\,C_i\,$, 
and $M_{K,0}$  stands for the lowest order kaon mass. The HKCHPT
Lagrangian thus reads:
\begin{eqnarray}
 {\cal L}_{\pi KK}^{(1)} &=& D_{\mu}K^{\dagger} D^{\mu}K-M_{K,0}^2 
K^{\dagger}K~,  \\
{\cal L}_{\pi KK}^{(2)} &=& A_1 {\rm Tr}\bigl(\Delta_{\mu}\Delta^{\mu}\bigr)
K^{\dagger}K+A_2 {\rm Tr}\bigl(\Delta^{\mu}\Delta^{\nu}\bigr) D_{\mu}
K^{\dagger}D_{\nu}K+A_3 K^{\dagger} \chi_+ K+A_4 {\rm Tr}
\bigl( \chi_+\bigr) K^{\dagger}K~,\\
{\cal L}_{\pi KK}^{(3)} &=& B_1 \Bigl( K^{\dagger}\bigl[\Delta^{\nu \mu},
\,\Delta_{\nu}\bigr] D_{\mu}K- D_{\mu}K^{\dagger}\bigl[\Delta^{\nu \mu},
\,\Delta_{\nu}\bigr]K\Bigr) +B_2 {\rm Tr}
\Bigl(\Delta^{\mu \nu}\Delta^{\rho}\Bigr)\Bigl( D_{\mu \nu}
K^{\dagger} D_{\rho}K+ D_{\rho}K^{\dagger} D_{\mu \nu}K \Bigr)
\nonumber \\ 
&+ &  B_3 \Bigl( K^{\dagger}\bigl[\Delta_{ \mu},\,\chi_-\bigr] 
D^{\mu}K- D_{\mu}K^{\dagger}\bigl[\Delta^{ \mu},\,\chi_-\bigr]K\Bigr)~,\\ 
{\cal L}_{\pi KK}^{(4)} &=& C_1 
 {\rm Tr}\Bigl(\Delta_{\nu}\Delta^{\mu \nu}\Bigr)\Bigl(K^{\dagger} 
D_{\mu}K+ D_{\mu}K^{\dagger}K \Bigr) + C_2  {\rm Tr}
\Bigl(\Delta^{\mu \rho}\Delta^{ \nu}\Bigr)\Bigl( D_{\mu \nu}K^{\dagger} 
D_{\rho}K+ D_{\rho}K^{\dagger} D_{\mu \nu}K \Bigr)\nonumber \\&+& C_3 
\Bigl( {\rm Tr}\bigl(\Delta^{\mu \nu}\Delta^{ \rho}\bigr)\bigl( D_{\mu \nu}
K^{\dagger} D_{\rho}K+ D_{\rho}K^{\dagger} D_{\mu \nu}K \bigr)
- 2\bigl(D^{\mu \nu}K^{\dagger}\Delta_{\mu}\Delta_{\nu \rho}
D^{\rho}K+D^{ \rho}K^{\dagger}\Delta_{\nu \rho}\Delta_{\mu }D^{\mu \nu}
K\bigr)\Bigr)\nonumber \\ &+& C_4  {\rm Tr}\Bigl(\Delta^{\mu \nu}
\Delta^{ \rho \sigma}\Bigr)\Bigl( D_{\mu \nu}K^{\dagger} D_{\rho \sigma}K
+ D_{\rho \sigma}K^{\dagger} D_{\mu \nu}K \Bigr) + C_5 
\Bigl(D_{\mu}K^{\dagger} \chi_+ D^{\mu}K-M_K^2 K^{\dagger} \chi_+ K \Bigr)
\nonumber \\ 
&+& C_6 \Bigl({\rm Tr}\bigl( \chi_+\bigr)D_{\mu}K^{\dagger} 
D^{\mu}K-M_K^2 {\rm Tr}\bigl( \chi_+\bigr) K^{\dagger}  K \Bigr)
+ C_7  {\rm Tr}\Bigl(\Delta_{\mu}\chi_-\Bigr)\Bigl( K^{\dagger} D_{\mu} 
 K+ D_{\mu}K^{\dagger}K \Bigr) \nonumber \\ &+& C_8  {\rm Tr}\Bigl(\Delta_{\mu}\Delta^{\mu }
\Bigr)K^{\dagger}\chi_+K + C_9  {\rm Tr}\Bigl(\Delta_{\mu}
\Delta^{\mu }\Bigr){\rm Tr}\Bigl( \chi_+\Bigr)K^{\dagger}K\nonumber\\
&+ & C_{10}  {\rm Tr}\Bigl(\Delta^{\mu }\Delta^{ \nu }\Bigr)\Bigl( D_{\mu }
K^{\dagger}\chi_+ D_{\nu}K+ D_{\nu}K^{\dagger}\chi_+ D_{\mu }K \Bigr)
\nonumber \\
&+& C_{11}  {\rm Tr}\Bigl(\Delta^{\mu }\Delta^{ \nu }\Bigr){\rm Tr}
\Bigl(\chi_+\Bigr)\Bigl( D_{\mu }K^{\dagger} D_{\nu}K+ D_{\nu}K^{\dagger} 
D_{\mu }K \Bigr) + C_{12} D_{\mu}K^{\dagger}\Bigl\{\bigl\{
\Delta^{\mu},\,\Delta^{\nu}\bigr\},\,\chi_+ \Bigr\} D_{\nu}K
\nonumber \\ 
&+& C_{13}{\rm Tr}\Bigl(\chi_+\Bigr)K^{\dagger} \chi_+ K + C_{14}{\rm Tr}
\Bigl(\chi_+^2\Bigr)K^{\dagger}  K +C_{15}\Bigl({\rm Tr}\bigl(\chi_+\bigr)
\Bigr)^2 K^{\dagger}  K +C_{16}{\rm Tr}\Bigl(\chi_-^2\Bigr) K^{\dagger}  K ~. 
\end{eqnarray}
From the power counting formula Eq.~(\ref{fpower}) it follows that loops 
start contributing to amplitudes at third  order. The infinities 
they generate are handled in the standard way, i.e. by renormalizing the 
LECs. Since ${\cal L}_{\pi KK}^{(2)}$ only accounts for second order 
tree contributions, the $A_i$ are finite. However, 
for reasons of notational consistency, we write $A_i^r$ in formulae 
describing renormalized observables, in analogy with $B_i^r$ and $C_i^r\,$.  

\def\theequation{\Alph{section}.\arabic{equation}}
\setcounter{equation}{0}
\section{Pion-kaon scattering amplitude in heavy kaon CHPT}
\label{app:heavyTpiK}

In this appendix, we present the pion-kaon scattering amplitude
$T^{3/2}_{\pi K \to \pi K}(\nu,t)$. As noted before, it does
not agree with the one given in \cite{Roessl} at various places.
To one--loop accuracy, it takes the form
\begin{eqnarray}
& & T^{\frac{3}{2}}_{\pi K \rightarrow \pi K}(\nu,t)=
  -\frac{1}{4 F_{\pi}^2}\nu
 -\frac{A_2^r}{16 F_{\pi}^2}\nu^2+
\frac{A_1^r}{2 F_{\pi}^2}t +(-A_1^r-2 A_3^r-4 A_4^r)
\frac{1}{F_{\pi}^2} M_{\pi}^2\nonumber\\& &
 -\frac{C_3^r}{16 F_{\pi}^2} \nu^3
-\frac{B_1^r}{4 F_{\pi}^2} \nu t +\Bigl(\frac{B_1^r}{2}
 - 2 B_3^r\Bigr)\frac{1}{F_{\pi}^2}\nu M_{\pi}^2
+\frac{1}{(4 \pi)^2}\frac{1}{F_{\pi}^4}
\Bigl(-\frac{1}{36}\nu t+\frac{1}{6} \nu M_{\pi}^2\Bigr)
\nonumber\\& & -\frac{C_4^r}{32 F_{\pi}^2} \nu^4
+\Bigl(-\frac{3 B_2^r}{16}+\frac{C_2^r}{16}
-\frac{C_3^r}{4}\Bigr)\frac{1}{F_{\pi}^2}\nu^2 t\nonumber\\ &&
+\Bigl(-\frac{A_2^r}{16} +\frac{C_1^r}{4}\Bigr)
\frac{1}{F_{\pi}^2}t^2+\Bigl(\frac{A_2^r C_5^r}{8}
+\frac{A_2^r C_6^r}{4} -\frac{C_{10}^r}{4} 
-\frac{C_{11}^r}{2}  -\frac{C_{12}^r}{4}\Bigr)
\frac{1}{F_{\pi}^2} \nu^2 M_{\pi}^2\nonumber\\& & 
+\Bigl(-A_1^r C_5^r-2 A_1^r C_6^r-\frac{C_1^r}{2}
+C_5^r+2 C_6^r+2 C_7^r+C_8^r+2 C_9^r\Bigr)
\frac{1}{F_{\pi}^2} t M_{\pi}^2\nonumber\\& & 
+\Bigl(2 A_1^r C_5^r+4 A_1^r C_6^r+4 A_3^r C_5^r
+8 A_3^r C_6^r+8 A_4^r C_5^r+16 A_4^r C_6^r
+\frac{4 A_3^r l_3^r}{F_{\pi}^2}+\frac{8 A_4^r l_3^r}
{F_{\pi}^2}-2 C_8^r-4 C_9^r \nonumber\\& &
-16 C_{13}^r-16 C_{14}^r
 -32 C_{15}^r-16 C_{16}^r\Bigr)
\frac{1}{F_{\pi}^2} M_{\pi}^4+\frac{1}{(4 \pi)^2}
\frac{1}{F_{\pi}^4}\Bigl(\frac{A_2^r M_K^2}{18} t^2
-\frac{13 A_2^r M_K^2}{36} t M_{\pi}^2
+\frac{A_2^r M_K^2}{6} M_{\pi}^4\Bigr)\nonumber\\& &  
+\biggl(\frac{1}{6} \nu M_{\pi}^2+\Bigl(-\frac{1}{8 M_K^2}
+\frac{A_2^r}{8}\Bigr)\nu^2 M_{\pi}^2+\Bigl(-\frac{1}{2}
+\frac{A_2^r M_K^2}{6} \Bigr)t M_{\pi}^2
\nonumber \\ &&
+\Bigl(1+\frac{3 A_1^r}{2}-\frac{A_2^r M_K^2}{12}+4 A_3^r 
+8 A_4^r\Bigr)M_{\pi}^4\biggr)\frac{1}{F_{\pi}^4}\mu_{\pi}
\nonumber\\& & +\biggl(\frac{1}{24} \nu t
-\frac{1}{6} \nu M_{\pi}^2+\Bigl(-\frac{A_1^r}{2}
-\frac{A_2^r M_K^2}{12}\Bigr)t^2+\Bigl(
\frac{5 A_1^r}{4}+\frac{3 A_2^r M_K^2}{8}+2 A_3^r 
+4 A_4^r \Bigr)t M_{\pi}^2\nonumber\\& & 
 +\Bigl(-\frac{A_1^r}{2}-\frac{A_2^r M_K^2}{6}-A_3^r
-2 A_4^r\Bigr) M_{\pi}^4\biggr)\frac{1}{F_{\pi}^4}
 J_{\pi \pi}^r(t) \\& & 
+\biggl(-\frac{1}{32 M_K}\nu^2+\Bigl(\frac{1}{64 M_K^3}
-\frac{A_1^r}{64 M_K^3}-\frac{A_2^r}{64M_K}\Bigr)\nu^3
+\frac{1}{16 M_K} \nu t   
+\Bigl(-\frac{1}{8 M_K}
-\frac{A_3^r}{2 M_K}-\frac{A_4^r}{M_K}\Bigr)\nu M_{\pi}^2
\biggr)\frac{1}{F_{\pi}^4}J_{\pi}^r (x_-)
\nonumber\\& & 
+\biggl(\frac{3}{32 M_K}\nu^2+\Bigl(\frac{3}{64 M_K^3}
+\frac{A_1^r}{64 M_K^3}+\frac{A_2^r}{64M_K}\Bigr)\nu^3
+\frac{3}{16 M_K} \nu t
+\Bigl(-\frac{3}{8 M_K}
+\frac{A_3^r}{2 M_K}+\frac{A_4^r}{M_K}\Bigr)\nu M_{\pi}^2
\biggr)\frac{1}{F_{\pi}^4}J_{\pi}^r (x_+)
\nonumber\\& &
 +\Bigl(-\frac{1}{512 M_K^4}\nu^4+\frac{1}{32 M_K^2}
 \nu^2 M_{\pi}^2 \Bigr)\frac{1}{F_{\pi}^4}G_{\pi}^r (x_-)
+\Bigl(-\frac{3}{512 M_K^4}\nu^4+\frac{3}{32 M_K^2}
 \nu^2 M_{\pi}^2 \Bigr)\frac{1}{F_{\pi}^4}G_{\pi}^r (x_+)
+{\cal O}(p^6)~, \nonumber
\end{eqnarray}
in terms of the loop integrals
\begin{eqnarray}
& &\int \frac{d^4 k}{(2 \pi)^4} \frac{i}{k^2-M_{\pi}^2}
\frac{1}{\omega-v\cdot k}=4 \omega L+J_{\pi}^r(\omega)~,\\
& & \int \frac{d^4 k}{(2 \pi)^4} \frac{i}{k^2-M_{\pi}^2}
\frac{1}{(\omega-v\cdot k)^2}=-4  L+G_{\pi}^r(\omega)~,
\end{eqnarray}
with $L \sim 1/(d-4)$ as usual in dimensional regularization
and $x_\pm = (\nu \pm t ) /4M_K$.
The loop function $J_{\pi \pi}^r(t)$ is taken from \cite{GL}
and $\mu_{\pi}$ is defined in Eq.~(\ref{mudef}). 
Note also that in Ref.~\cite{Roessl} matching relation were derived
by comparing this amplitude to the one obtained in standard SU(3)
CHPT \cite{BKMpik}. Most of these are correct, however, in two cases
we have found an error in the terms $\sim 1/\pi^2$ in the relations
for $B_1^r$ and $B_3^r$. The corrected matching conditions read:
\beqa
B_1^r &=& {1\over F^2} \left( -4L_3^r - {5 \over 576 \pi^2} - {5\over
108}{\arctan(\sqrt{2}) \over \sqrt{2}\pi^2} - {31\over 864} {\ln(4/3)
\over \pi^2} +{1\over 96} {\ln(\bar{M}_K^2 / \lambda^2) \over \pi^2}
+ {\cal O}(\bar{M}_K^2) \right)~, \\ 
B_3^r &=& {1\over F^2} \left( -L_3^r +L_5^r + {13 \over 2304 \pi^2} - {1\over
108}{\arctan(\sqrt{2}) \over \sqrt{2}\pi^2} + {7\over 1728} {\ln(4/3)
\over \pi^2} -{7\over 768} {\ln(\bar{M}_K^2 / \lambda^2) \over \pi^2}
+ {\cal O}(\bar{M}_K^2) \right)\,. 
\eeqa
The numerical analysis of the matching conditions derived from the
$\pi K$ amplitude leads to the numbers collected in table~\ref{coval}.

\def\theequation{\Alph{section}.\arabic{equation}}
\setcounter{equation}{0}
\section{Form factors and S-wave projected scattering amplitudes}
\label{app:forder}
In this appendix, we collect the one-loop representations of the various
non-strange scalar form factors and S-wave projected scattering amplitudes
appearing in Eq.~(\ref{impart}). The derivation of the scalar form factors 
$\Gamma_{\pi}$ and $\Gamma_{\eta}$  is completely analogous to the one of
$\Gamma_K$. One finds:
\begin{eqnarray}     
\Gamma_{\pi}(s)&=& M_{\pi}^2+\frac{M_{\pi}^2}{F^2}\Bigl[
L_4^r\bigl(-16M_{\pi}^2+8 s\bigr)+L_{5}^r\bigl(-8M_{\pi}^2+4s\bigr)+L_6^r 
32 M_{\pi}^2+L_8^r 16M_{\pi}^2\nonumber \\& &  +\bigl(\frac{1}{2}M_{\pi}^2-s\bigr)
 J^r_{\pi \pi}\bigl(s\bigr) -\frac{1}{4}s J^r_{KK}\bigl(s\bigr) - 
\frac{1 }{18}M_{\pi}^2 J^r_{\eta \eta}\bigl(s\bigr)\Bigr]+{\cal O}(p^6) ~,
\\ \nonumber 
\\  
\Gamma_{\eta}(s)&=& \frac{M_{\pi}^2}{3}+\frac{M_{\pi}^2}{F^2}
\Bigl[L_4^r\bigl(\frac{16}{3}M_{\pi}^2-\frac{64}{3}M_{K}^2+8 s\bigr)+L_{5}^r
\bigl(\frac{40}{9}M_{\pi}^2-\frac{64}{9}M_{K}^2+\frac{4}{3}s\bigr)\nonumber\\
& & +L_6^r \bigl(-\frac{32}{3} M_{\pi}^2+\frac{128}{3}M_{K}^2\bigr) +L_7^r 
\bigl(\frac{128}{3}M_{\pi}^2-\frac{128}{3}M_{K}^2\bigr) +L_8^r\frac{ 16}{3}M_{\pi}^2 
-\frac{2}{3}M_{\pi}^2 \mu_{\pi} +\frac{2}{3}M_{K}^2 \mu_{K}\nonumber \\
&&-\frac{1}{2} M_{\pi}^2  J^r_{\pi \pi}\bigl(s\bigr) 
+\bigl(\frac{2}{3} M_K^2-\frac{3}{4}s\bigr) 
J^r_{KK}\bigl(s\bigr) 
+\bigl( \frac{7 }{54}M_{\pi}^2-\frac{8}{27} 
M_K^2\bigr) J^r_{\eta \eta}\bigl(s\bigr)\Bigr]+{\cal O}(p^6) ~,
\end{eqnarray}
in terms of the physical meson masses and we set $F = F_\pi$ throughout.
Next, we consider the various isospin zero  
$KK \to$2~Goldstone bosons scattering amplitudes.
From these, we consider the projection on the  $l=0$ components (S-waves) using
(for generic Goldstone bosons $\Phi_a$)
\begin{equation}
t^{0}_{0,\Phi_a \Phi_c \rightarrow \Phi_b \Phi_d}(s)= \frac{1}{64 \pi}\int_{-1}^1  
dz T^{0}_{\Phi_a \Phi_c \rightarrow \Phi_b \Phi_d}(s, t, u)~,
\label{proint}
\end{equation} 
with $z = \cos (\theta)$ and  the  angular dependence is implicitly contained in 
$t$ and $u\,$. The pertinent Mandelstam variables are defined by the
following kinematics
\begin{equation}
K(p_1) + K(p_3)\rightarrow \Phi_a (p_2) + \Phi_a(p_4)~,
\end{equation}
where $\Phi_a$ stands for pion, kaon, or eta degrees of freedom. 
The coordinate frame is chosen such that $\theta$ is the angle included by 
$\vec{p}_1$ and $\vec{p}_2\,$. 
The results for the three amplitudes pertinent to our case will now be given.

\smallskip
\noindent
$\bullet\,\,KK \rightarrow \pi \pi$ scattering
   
\noindent
The Mandelstam variables for this configuration are
\begin{eqnarray}
s &=& 4\bigl(\vec{p}_1\,^2+M_K^2 \bigr)~,\nonumber \\ 
t &=& -2 \vec{p}_1\,^2 + M_{\pi}^2-M_K^2 + 2 |\vec{p}_1| \sqrt{ \vec{p}_1\,^2
-M_{\pi}^2+M_K^2}\, z~,\nonumber\\
u &=&  -2 \vec{p}_1\,^2 + M_{\pi}^2-M_K^2 - 2 |\vec{p}_1| \sqrt{ \vec{p}_1\,^2
-M_{\pi}^2+M_K^2}\, z~.
\end{eqnarray}
The $I=0,\,l=0$ partial amplitude is then given by the following expression:
\begin{eqnarray}
& &  t^{0}_{0,K K \rightarrow \pi \pi}(s)
=-\frac{1}{64 \pi} \sqrt{\frac{3}{2}}\biggl\{\frac{s}{F_{\pi}^2}\nonumber\\ 
& &
+\frac{1}{F^4}\Bigl(L_1^r \bigl(128 M_{\pi}^2 M_K^2-64 M_{\pi}^2s-64 M_K^2s+32 s^2\bigr)
+L_2^r \bigl(\frac{128}{3} M_{\pi}^2 M_K^2-\frac{32}{3} M_{\pi}^2s\nonumber\\
& & \quad\quad-\frac{32}{3} M_K^2s+\frac{32}{3} s^2\bigr)+L_3^r \bigl(\frac{128}{3}
 M_{\pi}^2 M_K^2-\frac{56}{3} M_{\pi}^2s-\frac{56}{3} M_K^2s
+\frac{32}{3} s^2\bigr)\nonumber\\
& & \quad\quad+L_4^r \bigl(-128 M_{\pi}^2 M_K^2+32 M_{\pi}^2s+32 M_K^2s\bigr)
+L_5^r\bigl(-32 M_{\pi}^2 M_K^2+8 M_{\pi}^2s\bigr)\nonumber\\
& & \quad\quad+L_6^r 128 M_{\pi}^2 M_K^2+L_8^r 64 M_{\pi}^2 M_K^2\nonumber\\
& & \quad\quad+ \bigl(\frac{1}{2} M_{\pi}^4-\frac{13}{6} M_{\pi}^2 M_K^2
-\frac{1}{2} M_K^4+\frac{19}{24} M_{\pi}^2s+\frac{7}{24} M_K^2s-\frac{65}{48} 
s^2\bigr) \mu_{\pi}\nonumber\\& & \quad\quad+ \bigl(-\frac{11}{27} M_{\pi}^4
-\frac{8}{3} M_{\pi}^2 M_K^2+\frac{20}{27} M_K^4+\frac{1}{4} M_{\pi}^2s+\frac{5}{4}
 M_K^2s-\frac{9}{8} s^2\bigr) \mu_{K}\nonumber\\& & \quad\quad+ \bigl(-\frac{5}{54}
 M_{\pi}^4+\frac{11}{18} M_{\pi}^2 M_K^2-\frac{13}{54} M_K^4-\frac{13}{24} M_{\pi}^2s
+\frac{11}{24} M_K^2s-\frac{1}{48} s^2\bigr) \mu_{\eta}\nonumber\\& & \quad\quad+
\bigl(\frac{1}{2} M_{\pi}^2s-s^2\bigr)\tilde{J}^r_{\pi \pi}\bigl(s\bigr)-\frac{3}{4}s^2 
\tilde{J}^r_{KK}\bigl(s\bigr)+\bigl(\frac{4}{9}M_{\pi}^2 M_K^2-\frac{1}{2}M_{\pi}^2s\bigr)
\tilde{J}^r_{\eta \eta}\bigl(s\bigr)\nonumber\\& & \quad\quad+\frac{1}{(4 \pi)^2} 
\bigl(-\frac{1}{18}M_{\pi}^4+\frac{59}{9} M_{\pi}^2 M_K^2+\frac{23}{18} M_{K}^4-
\frac{53}{36} M_{\pi}^2s-\frac{125}{36} M_K^2s+\frac{43}{18} s^2\bigr)\nonumber\\
& &\quad\quad +\frac{1}{\sqrt{s-4 M_{\pi}^2}\sqrt{s-4 M_{K}^2}} \ln 
\Bigl(\frac{2M_{\pi}^2+2 M_K^2-s- \sqrt{s-4 M_{\pi}^2}\sqrt{s-4 M_{K}^2}}{2M_{\pi}^2
+2 M_K^2-s+ \sqrt{s-4 M_{\pi}^2}\sqrt{s-4 M_{K}^2}}\Bigr)\nonumber\\ 
& & \quad\quad\Bigl(\bigl(\frac{1}{2} M_{\pi}^6+2 M_{\pi}^4 M_K^2-2 M_{\pi}^2 M_K^4
-\frac{1}{2} M_K^6-\frac{3}{8} M_{\pi}^4s+\frac{3}{8} M_K^4s\bigr) \mu_{\pi}
+\bigl(-\frac{11}{27} M_{\pi}^6\nonumber\\& & \quad\quad-\frac{7}{3} M_{\pi}^4 M_K^2
+2 M_{\pi}^2 M_K^4+\frac{20}{27} M_K^6+\frac{5}{12} M_{\pi}^4s-\frac{1}{3} M_{\pi}^2
 M_K^2s-\frac{1}{12} M_K^4s\bigr) \mu_{K}\nonumber\\& & \quad\quad+\bigl(-\frac{5}{54} 
M_{\pi}^6+\frac{1}{3} M_{\pi}^4 M_K^2-\frac{13}{54} M_K^6-\frac{1}{24} M_{\pi}^4s
+\frac{1}{3} M_{\pi}^2 M_K^2s-\frac{7}{24} M_K^4s\bigr) \mu_{\eta}\nonumber\\
& & \quad\quad+\frac{1}{(4\pi)^2}\bigl(-\frac{1}{18} M_{\pi}^6+\frac{25}{18} 
M_{\pi}^4 M_K^2-\frac{47}{18} M_{\pi}^2 M_K^4+\frac{23}{18} M_K^6-\frac{5}{18} 
M_{\pi}^4s\nonumber\\& & \quad\quad+\frac{5}{9}M_{\pi}^2 M_K^2s-\frac{5}{18} 
M_K^4s\bigr)\Bigr)\nonumber \\ & &\quad\quad +\int_{-1}^1 dz \Bigl(  
\bigl(-\frac{1}{8}\frac{M_{\pi}^8}{t^2}-\frac{1}{4}M_{\pi}^4+\frac{1}{2}
\frac{M_{\pi}^6 M_K^2}{t^2}-\frac{3}{4}\frac{ M_{\pi}^4 M_K^4}{t^2}-\frac{3}{2} 
M_{\pi}^2 M_K^2\nonumber \\ & &\quad\quad \quad+\frac{1}{2}\frac{M_{\pi}^2 M_K^6}{ t^2}
-\frac{1}{8}\frac{M_K^8}{t^2} -\frac{1}{4} M_K^4-\frac{1}{16} \frac{M_{\pi}^4s}{t}
+\frac{1}{8}M_{\pi}^2s+\frac{1}{8} \frac{M_{\pi}^2 M_K^2 s}{t}\nonumber\\
& & \quad\quad\quad-\frac{1}{16}\frac{M_K^4s}{t}+\frac{1}{8}M_K^2s-\frac{1}{16}st
+M_{\pi}^2t +M_K^2t-\frac{5}{8}t^2\bigr)   \tilde{J}^r_{\pi K}\bigl(t\bigr)\nonumber\\
& &\quad\quad \quad
+\bigl(-\frac{1}{72}\frac{M_{\pi}^8}{t^2}-\frac{1}{18}\frac{M_{\pi}^6}{t}+\frac{1}{36}
M_{\pi}^4+\frac{1}{18}\frac{M_{\pi}^6 M_K^2}{ t^2}+\frac{2}{9}\frac{M_{\pi}^4 M_K^2}{t}
-\frac{1}{12}\frac{M_{\pi}^4 M_K^4}{ t^2}\nonumber \\ 
& &\quad\quad \quad-\frac{7}{18} 
M_{\pi}^2 M_K^2 -\frac{5}{18}\frac{ M_{\pi}^2 M_K^4}{t}+\frac{1}{18}
\frac{M_{\pi}^2 M_K^6}{t^2}-\frac{1}{72}\frac{M_K^8}{ t^2}+\frac{1}{9}\frac{M_K^6}{t}
-\frac{11}{36} M_K^4\nonumber \\ & &\quad\quad \quad-\frac{1}{144} \frac{M_{\pi}^4s}{t}
-\frac{1}{24} M_{\pi}^2s+\frac{1}{72} \frac{M_{\pi}^2 M_K^2s}{t}-\frac{1}{144} 
\frac{M_K^4s}{t}+\frac{7}{24} M_K^2s -\frac{1}{16}s t\nonumber \\ 
& & \quad \quad\quad+\frac{1}{6} M_{\pi}^2 t+\frac{1}{3} M_K^2t-\frac{1}{8}t^2 \bigr)
\tilde{J}^r_{K \eta}\bigl(t\bigr)\nonumber\\ 
& &\quad \quad\quad+
\bigl(-\frac{1}{8}\frac{M_{\pi}^8}{u^2}-\frac{3}{4}M_{\pi}^4+\frac{1}{2}
\frac{M_{\pi}^6 M_K^2}{u^2}-\frac{3}{4}\frac{ M_{\pi}^4 M_K^4}{u^2}-\frac{5}{2} 
M_{\pi}^2 M_K^2+\frac{1}{2}\frac{M_{\pi}^2 M_K^6}{ u^2}\nonumber \\ 
& &\quad\quad \quad-\frac{1}{8}\frac{M_K^8}{u^2} -\frac{3}{4} M_K^4-\frac{1}{16} 
\frac{M_{\pi}^4s}{u}+\frac{3}{2}M_{\pi}^2s+\frac{1}{8} \frac{M_{\pi}^2 M_K^2 s}{u}
-\frac{1}{16}\frac{M_K^4s}{u}\nonumber\\& & \quad\quad\quad+\frac{3}{2}M_K^2s
-\frac{9}{16}s^2-\frac{19}{16}st+\frac{3}{2}M_{\pi}^2t +\frac{3}{2}M_K^2t
-\frac{5}{8}t^2\bigr)   \tilde{J}^r_{\pi K}\bigl(u\bigr)\nonumber\\
& &\quad\quad \quad
+\bigl(-\frac{1}{72}\frac{M_{\pi}^8}{u^2}-\frac{1}{18}\frac{M_{\pi}^6}{u}
-\frac{5}{36}M_{\pi}^4+\frac{1}{18}\frac{M_{\pi}^6 M_K^2}{ u^2}+\frac{2}{9}
\frac{M_{\pi}^4 M_K^2}{u}-\frac{1}{12}\frac{M_{\pi}^4 M_K^4}{ u^2}\nonumber \\ 
& &\quad\quad \quad-\frac{7}{18} M_{\pi}^2 M_K^2 -\frac{5}{18}\frac{ M_{\pi}^2 M_K^4}{u}
+\frac{1}{18}\frac{M_{\pi}^2 M_K^6}{u^2}-\frac{1}{72}\frac{M_K^8}{ u^2}+\frac{1}{9}
\frac{M_K^6}{u}-\frac{5}{36} M_K^4\nonumber \\ & &\quad\quad \quad-\frac{1}{144} 
\frac{M_{\pi}^4s}{u}+\frac{1}{6} M_{\pi}^2s+\frac{1}{72} \frac{M_{\pi}^2 M_K^2s}{u}
-\frac{1}{144} \frac{M_K^4s}{u}+\frac{1}{3} M_K^2s -\frac{1}{16}s^2\nonumber\\
& & \quad\quad\quad-\frac{3}{16}s t+\frac{1}{3} M_{\pi}^2 t+\frac{1}{6} M_K^2t
-\frac{1}{8}t^2 \bigr)\tilde{J}^r_{K \eta}\bigl(u\bigr)\Bigl)\Bigl)\biggl\}+
{\cal O}(p^6)~.\label{hoch}
\end{eqnarray}

\medskip
\noindent
$\bullet\,\,KK \rightarrow K K$ scattering 

\noindent  
The Mandelstam variables for this configuration are
\beq
s = 4\bigl(\vec{p}_1\,^2+M_K^2 \bigr)~,\quad
t = 2 \vec{p}_1\,^2 \bigl(z-1\bigr)~,\quad
u =  -2 \vec{p}_1\,^2 \bigl(z+1\bigr)~.
\eeq
In this case, besides the $t$--channel and the $u$--channel loop functions, 
the terms proportional to $1/t$ cannot be integrated analytically.
The $I=0,\,l=0$ partial amplitude is then given by the following expression:
\begin{eqnarray}
& &  t^{0}_{0,K K \rightarrow K K}(s)=\frac{1}{64 \pi} 
\biggl\{\frac{3s}{2 F_{\pi}^2}\nonumber\\ & &
+\frac{1}{F^4}\Bigl(L_1^r\bigl(\frac{448}{3} M_K^4-\frac{416}{3} M_K^2s+\frac{112}{3}s^2 
\bigr)+L_2^r\bigl(\frac{256}{3} M_K^4-\frac{176}{3} M_K^2s+\frac{64}{3}s^2 \bigr)
\nonumber\\ & & \quad\quad+L_3^r\bigl(64 M_K^4-56 M_K^2s+16 s^2\bigr)
+L_4^r \bigl(-128 M_K^4+48 M_K^2s\bigr)\nonumber\\& & \quad\quad+L_5^r 
\bigl(-48 M_K^4+12 M_{\pi}^2s\bigr)+L_6^r192 M_K^4+L_8^r 96 M_K^4\nonumber\\
& & \quad\quad+
 \bigl( \frac{1}{2}M_{\pi}^2 M_K^2-\frac{13}{6} M_K^4-\frac{3}{2} M_{\pi}^2s
+\frac{11}{24} M_K^2s-\frac{13}{24} s^2\bigr) \mu_{\pi}\nonumber\\
& & \quad\quad+ \bigl(-\frac{16}{3} M_K^4+\frac{35}{12} M_K^2s-\frac{41}{24} s^2\bigr) 
\mu_{K}+ \bigl(-\frac{1}{2} M_{\pi}^2 M_K^2-\frac{17}{6} M_K^4-\frac{3}{4} M_{\pi}^2s
+\frac{45}{8} M_K^2s\nonumber\\& & \quad\quad-\frac{3}{2} s^2\bigr) \mu_{\eta}
-\frac{3}{8}s^2 \tilde{J}^r_{\pi \pi}\bigl(s\bigr)-\frac{9}{8}s^2 
\tilde{J}^r_{KK}\bigl(s\bigr)
+\bigl(-\frac{8}{9} M_K^4+2 M_K^2s-\frac{9}{8}s^2\bigr) \tilde{J}^r_{\eta \eta}
\bigl(s\bigr)\nonumber\\& & \quad\quad+\frac{1}{(4\pi)^2}\bigl(2 M_{\pi}^2 M_K^2
+\frac{41}{3} M_K^4-\frac{3}{2} M_{\pi}^2s-\frac{107}{12} M_K^2s+\frac{43}{12} 
s^2\bigr)\nonumber \\ & & \quad\quad+\int_{-1}^1 dz \Bigl(\bigl(-\frac{2}{3} 
\frac{M_{\pi}^2 M_K^4}{t}+\frac{2}{3} \frac{M_K^6}{t}\bigr)\mu_{\pi}+ 
\bigl(\frac{2}{3} \frac{M_{\pi}^2 M_K^4}{t}-\frac{2}{3} \frac{M_K^6}{t}\bigr)
\mu_{\eta}\nonumber\\ & &\quad \quad\quad+ \bigl(- M_{\pi}^2 M_K^2+\frac{1}{2} 
M_{\pi}^2s-\frac{1}{8}st +\frac{1}{4} M_{\pi}^2t+\frac{1}{4} M_K^2t-\frac{5}{32} 
t^2\bigr)\tilde{J}^r_{\pi \pi}\bigl(t\bigr)\nonumber\\
& & \quad\quad\quad+ 
\bigl(-2 M_{K}^4 +M_{K}^2s-\frac{1}{4}st + M_{K}^2t-\frac{1}{2} t^2\bigr)
\tilde{J}^r_{KK}\bigl(t\bigr)+\bigl(-\frac{2}{9} M_K^4+\frac{1}{2} M_K^2t\nonumber\\
& & \quad\quad\quad-\frac{9}{32} t^2 \bigl)\tilde{J}^r_{\eta\eta}\bigl(t\bigr)   
+\bigl(- M_K^4+\frac{3}{2} M_K^2t-\frac{9}{16} t^2 \bigl)\tilde{J}^r_{\pi\eta}
\bigl(t\bigr) \nonumber\\ 
& & \quad\quad\quad                        
+ \bigl(-3  M_K^4+3 M_{K}^2s-\frac{3}{4}s^2-\frac{3}{2}st +3 M_K^2t
-\frac{3}{4} t^2\bigr)\tilde{J}^r_{KK}\bigl(u\bigr)\Bigr)\Bigr)\biggr\}
+{\cal O}(p^6)~.
\end{eqnarray}

\medskip
\noindent
$\bullet\,\,KK \rightarrow \eta\eta$ scattering 

\noindent  
The Mandelstam variables for this case are
\begin{eqnarray}
s &=& 4\bigl(\vec{p}_1\,^2+M_K^2 \bigr)~,\nonumber \\ 
t &=& -2 \vec{p}_1\,^2-\frac{1}{3} M_{\pi}^2+\frac{1}{3} M_K^2
+ 2 |\vec{p}_1| \sqrt{ \vec{p}_1\,^2+\frac{1}{3}M_{\pi}^2-\frac{1}{3}M_K^2}\, 
z~,\nonumber\\
u &=&  -2 \vec{p}_1\,^2-\frac{1}{3} M_{\pi}^2+\frac{1}{3} M_K^2- 2 |\vec{p}_1|
 \sqrt{ \vec{p}_1\,^2+\frac{1}{3}M_{\pi}^2-\frac{1}{3}M_K^2}\, z ~.
\end{eqnarray}
The $I=0,\,l=0$ partial amplitude is then given by the following expression:
\begin{eqnarray}
& &  t^{0}_{0,K K \rightarrow \eta\eta}(s)
=\frac{1}{64 \pi} \sqrt{2} \biggl\{\frac{1}{2 F_{\pi}^2}\Bigl(-\frac{4}{3} M_K^2+\frac{3}{2}s\Bigr)\nonumber\\ & &
+\frac{1}{F^4}\Bigl(L_1^r \bigl(-\frac{64}{3} M_{\pi}^2 M_K^2+\frac{256}{3}
 M_K^4+\frac{32}{3} M_{\pi}^2s-\frac{224}{3} M_K^2s+16 s^2\bigr)\nonumber\\
& & \quad\quad+L_2^r \bigl(-\frac{64}{9} M_{\pi}^2 M_K^2+\frac{256}{9} M_K^4
+\frac{16}{9} M_{\pi}^2s-\frac{112}{9} M_K^2s+\frac{16}{3} s^2\bigr)\nonumber\\
& & \quad\quad+L_3^r \bigl(-\frac{256}{27} M_{\pi}^2 M_K^2+\frac{1024}{27} M_K^4
+\frac{124}{27} M_{\pi}^2s-\frac{868}{27} M_K^2s+\frac{64}{9} s^2\bigr)\nonumber\\
& & \quad\quad+L_4^r \bigl(\frac{64}{3} M_{\pi}^2 M_K^2-\frac{256}{3 }M_K^4
-\frac{16}{3} M_{\pi}^2s+\frac{112}{3} M_K^2s\bigr)\nonumber\\& & \quad\quad
+L_5^r\bigl(-\frac{16}{9}M_{\pi}^2 M_K^2-\frac{224}{9} M_K^4+12 M_{\pi}^2s\bigr)
+L_6^r\bigl( -\frac{64}{3} M_{\pi}^2 M_K^2+\frac{256}{3} M_K^4\bigr)\nonumber\\
& & \quad\quad+L_7^r\bigl( -\frac{128}{3} M_{\pi}^2 M_K^2+\frac{128}{3} M_K^4\bigr)
+L_8^r\bigl( -32 M_{\pi}^2 M_K^2+64 M_K^4\bigr)\nonumber\\& & \quad\quad+ 
\bigl(\frac{3}{4} M_{\pi}^4-\frac{17}{12} M_{\pi}^2 M_K^2+\frac{37}{12} M_K^4
-\frac{155}{48} M_{\pi}^2s+\frac{29}{48} M_K^2s-\frac{1}{32} s^2\bigr) \mu_{\pi}\nonumber\\
& & \quad\quad+ \bigl(-\frac{7}{6} M_{\pi}^4+\frac{23}{3} M_{\pi}^2 M_K^2-\frac{86}{9} 
M_K^4+\frac{1}{24} M_{\pi}^2s+\frac{65}{24} M_K^2s-\frac{19}{16} s^2\bigr) \mu_{K}
\nonumber\\& & 
\quad\quad+ \bigl(\frac{5}{12} M_{\pi}^4-\frac{335}{108} M_{\pi}^2 M_K^2+\frac{347}{108}
 M_K^4+\frac{29}{48} M_{\pi}^2s-\frac{47}{48} M_K^2s-\frac{1}{32} s^2\bigr) 
\mu_{\eta}\nonumber\\ & & \quad\quad-\frac{1}{4} M_{\pi}^2s \tilde{J}^r_{\pi \pi}\
bigl(s\bigr)+\bigl(M_K^2s-\frac{9}{8}s^2\bigr)\tilde{J}^r_{KK}\bigl(s\bigr)
+\bigl(-\frac{14}{27} M_{\pi}^2 M_K^2+\frac{32}{27}M_K^4\nonumber\\
& & \quad\quad+\frac{7}{12} M_{\pi}^2s-\frac{4}{3} M_K^2s\bigr)\tilde{J}^r_{\eta\eta}
\bigl(s\bigr)\nonumber\\& & \quad\quad+\frac{1}{(4\pi)^2} \bigl(\frac{5}{4} M_{\pi}^4
-\frac{149}{54} M_{\pi}^2 M_K^2+\frac{751}{108} M_K^4-\frac{67}{72} M_{\pi}^2s
-\frac{179}{72} M_K^2s
+\frac{13}{12} s^2\bigr)\nonumber\\& & \quad\quad+\frac{1}{\sqrt{3}\sqrt{3s+4 M_{\pi}^2
-16 M_K^2}\sqrt{s-4 M_{K}^2}}\nonumber\\ & & \quad\quad \ln \Bigl(\frac{-2M_{\pi}^2+14
 M_K^2-3s-\sqrt{3} \sqrt{3s+4 M_{\pi}^2-16M_K^2}\sqrt{s-4 M_{K}^2}}{-2M_{\pi}^2+14 
M_K^2-3s+\sqrt{3} \sqrt{3s+4 M_{\pi}^2-16 M_K^2}\sqrt{s-4 M_{K}^2}}\Bigr)\nonumber\\
& & \quad\quad\Bigl(\bigl(-\frac{3}{4} M_{\pi}^6+\frac{9}{2} M_{\pi}^4 M_K^2
-\frac{15}{4} M_K^6-\frac{27}{16} M_{\pi}^4s+\frac{27}{16} M_K^4s\bigr) \mu_{\pi}
\nonumber\\& & 
\quad\quad+\bigl(\frac{7}{6} M_{\pi}^6-\frac{17}{2} M_{\pi}^4 M_K^2
+\frac{116}{9} M_{\pi}^2 M_K^4-\frac{50}{9} M_K^6+\frac{15}{8} M_{\pi}^4s-
\frac{3}{2} M_{\pi}^2 M_K^2s\nonumber\\& & \quad\quad-\frac{3}{8} M_K^4s\bigr) 
\mu_{K}+\bigl(-\frac{5}{12} M_{\pi}^6+4 M_{\pi}^4 M_K^2-\frac{116}{9}M_{\pi}^2M_K^4
+\frac{335}{36} M_K^6-\frac{3}{16} M_{\pi}^4s\nonumber\\& & \quad\quad+\frac{3}{2} 
M_{\pi}^2 M_K^2s-\frac{21}{16} M_K^4s\bigr) \mu_{\eta}+\frac{1}{(4\pi)^2}
\bigl(-\frac{5}{4} M_{\pi}^6+\frac{79}{12} M_{\pi}^4 M_K^2-\frac{113}{12} 
M_{\pi}^2 M_K^4\nonumber\\& & \quad\quad+\frac{49}{12} M_K^6
-\frac{5}{4} M_{\pi}^4s+\frac{5}{2}M_{\pi}^2 M_K^2s-\frac{5}{4} M_K^4s\bigr)\Bigr)
\nonumber \\
 & &\quad\quad +\int_{-1}^1 dz \Bigl(\bigl(-\frac{1}{48} \frac{M_{\pi}^8}{t^2}
-\frac{1}{12}\frac{M_{\pi}^6}{t}+\frac{1}{24}M_{\pi}^4+\frac{1}{12}\frac{ M_{\pi}^6 
M_K^2}{ t^2}+\frac{1}{3} \frac{M_{\pi}^4 M_K^2}{t}-\frac{1}{8} 
\frac{M_{\pi}^4 M_K^4}{t^2}\nonumber \\ 
& &\quad \quad\quad -\frac{7}{12}M_{\pi}^2 M_K^2 -\frac{5}{12}\frac{M_{\pi}^2 M_K^4}{t}
+\frac{1}{12}\frac{ M_{\pi}^2 M_K^6}{ t^2}
-\frac{1}{48}\frac{M_K^8}{t^2}+\frac{1}{6}\frac{M_K^6}{ t}-\frac{11}{24}M_K^4\nonumber \\ 
& &\quad \quad\quad -\frac{3}{32} \frac{M_{\pi}^4 s}{t}+\frac{3}{16} M_{\pi}^2s
+\frac{3}{16} \frac{M_{\pi}^2 M_K^2 s}{t}+\frac{3}{16} M_K^2s-\frac{3}{32}
\frac{ M_K^4 s}{t}-\frac{3}{32} st\nonumber\\ & & \quad\quad\quad+\frac{1}{4} M_{\pi}^2t
+\frac{1}{2} M_K^2t-\frac{3}{16} t^2 \bigr)\tilde{J}^r_{\pi, K}\bigl(t\bigr)
+\bigl(-\frac{1}{432} \frac{M_{\pi}^8}{t^2}-\frac{1}{36}\frac{M_{\pi}^6}{t}
-\frac{1}{8}M_{\pi}^4\nonumber\\& & \quad\quad\quad+\frac{1}{108}
\frac{ M_{\pi}^6 M_K^2}{ t^2}+\frac{7}{36} \frac{M_{\pi}^4 M_K^2}{t}-\frac{1}{72} 
\frac{M_{\pi}^4 M_K^4}{t^2}+\frac{11}{12}M_{\pi}^2 M_K^2 -\frac{11}{36}\frac{M_{\pi}^2 M_K^4}{t}\nonumber\\
& & \quad\quad\quad+\frac{1}{108}\frac{ M_{\pi}^2 M_K^6}{ t^2}
-\frac{1}{432}\frac{M_K^8}{t^2}+\frac{5}{36}\frac{M_K^6}{ t}-\frac{161}{72}M_K^4
-\frac{1}{96} \frac{M_{\pi}^4 s}{t}-\frac{1}{16} M_{\pi}^2s\nonumber\\
& & \quad\quad\quad+\frac{1}{48} \frac{M_{\pi}^2 M_K^2 s}{t}+\frac{7}{16} 
M_K^2s-\frac{1}{96}\frac{ M_K^4 s}{t}-\frac{3}{32} st-\frac{1}{4} M_{\pi}^2t+\frac{5}{4} M_K^2t-\frac{3}{16} t^2 \bigr)\tilde{J}^r_{ K \eta}\bigl(t\bigr)\nonumber\\
& & \quad\quad\quad
+\bigl(-\frac{1}{48} \frac{M_{\pi}^8}{u^2}-\frac{1}{12}\frac{M_{\pi}^6}{u}
-\frac{5}{24}M_{\pi}^4+\frac{1}{12}\frac{ M_{\pi}^6 M_K^2}{ u^2}+\frac{1}{3} 
\frac{M_{\pi}^4 M_K^2}{u}-\frac{1}{8} \frac{M_{\pi}^4 M_K^4}{u^2}\nonumber \\ 
& & \quad\quad \quad+\frac{17}{12}M_{\pi}^2 M_K^2 -\frac{5}{12}\frac{M_{\pi}^2 
M_K^4}{u}+\frac{1}{12}\frac{ M_{\pi}^2 M_K^6}{ u^2}
-\frac{1}{48}\frac{M_K^8}{u^2}+\frac{1}{6}\frac{M_K^6}{ u}-\frac{53}{24}M_K^4\nonumber \\ 
& & \quad\quad \quad-\frac{3}{32} \frac{M_{\pi}^4 s}{u}-\frac{1}{4} M_{\pi}^2s
+\frac{3}{16} \frac{M_{\pi}^2 M_K^2 s}{u}+ M_K^2s-\frac{3}{32}\frac{ M_K^4 s}{u}
-\frac{9}{32} st-\frac{1}{2} M_{\pi}^2t\nonumber\\
& & \quad\quad\quad+\frac{5}{4} M_K^2t-\frac{3}{16} t^2 \bigr)\tilde{J}^r_{\pi, K}
\bigl(u\bigr)
+\bigl(-\frac{1}{432} \frac{M_{\pi}^8}{u^2}-\frac{1}{36}\frac{M_{\pi}^6}{u}
-\frac{1}{24}M_{\pi}^4+\frac{1}{108}\frac{ M_{\pi}^6 M_K^2}{ u^2}\nonumber\\
& &\quad \quad\quad+\frac{7}{36} \frac{M_{\pi}^4 M_K^2}{u}-\frac{1}{72} 
\frac{M_{\pi}^4 M_K^4}{u^2}+\frac{1}{12}M_{\pi}^2 M_K^2 -\frac{11}{36}\frac{M_{\pi}^2 M_K^4}{u}+\frac{1}{108}\frac{ M_{\pi}^2 M_K^6}{ u^2}
\nonumber\\& & \quad\quad\quad-\frac{1}{432}\frac{M_K^8}{u^2}+\frac{5}{36}
\frac{M_K^6}{ u}-\frac{35}{72}M_K^4 -\frac{1}{96} \frac{M_{\pi}^4 s}{u}
+\frac{1}{48} \frac{M_{\pi}^2 M_K^2 s}{u}+\frac{1}{2} M_K^2s\nonumber\\
& & \quad\quad\quad-\frac{1}{96}\frac{ M_K^4 s}{u}-\frac{3}{32}s^2-\frac{9}{32} 
st+\frac{1}{2} M_K^2t-\frac{3}{16} t^2 \bigr)\tilde{J}^r_{ K \eta}
\bigl(u\bigr)\Bigr)\Bigr)\biggr\}+{\cal O}(p^6)~.
\end{eqnarray}
Throughout, we have employed the modified loop functions
 \begin{eqnarray}
\tilde{J}_{aa}(t)&=& J_{aa}(t)+\frac{1}{(4\pi)^2}-\mu_a~,\\
\tilde{J}_{ab}(t)&=& J_{ab}(t)+\frac{1}{(4\pi)^2}
+\frac{M_a^2-M_b^2}{2t}\bigl(\mu_b-\mu_a\bigr)
-\frac{1}{2}\bigl(\mu_b+\mu_a\bigr)~,
\label{modloop}
\end{eqnarray}
for $a,\,b\in \{\pi,\,K,\,\eta\}\,$.

\bigskip 

\end{document}